\documentclass[a4paper, 12pt]{article} % Font size (can be 10pt, 11pt or 12pt) and paper size (remove a4paper for US letter paper)
\usepackage[right=1.1in,left=1.1in,top=1.1in,bottom=1.1in]{geometry}
\usepackage{graphicx} 
\usepackage{braket}
\usepackage{mathrsfs}
\usepackage[T1]{fontenc} % Required for accented characters
\usepackage{amsmath}
\usepackage{pxfonts}
\usepackage[T1]{fontenc}
\usepackage{amssymb}
\usepackage{natbib}
\usepackage{epstopdf}
\usepackage{setspace}
\usepackage{enumitem}
\usepackage{sectsty}
\usepackage{wrapfig} % Allows in-line images
\sectionfont{\large}
\bibpunct{(}{)}{,}{a}{}{,}
\usepackage{tikz}   %TikZ is required for this to work.  Make sure this exists before the next line
\usepackage{tikz-3dplot} %requires 3dplot.sty to be in same directory, or in your LaTeX installation
\usepackage[protrusion=true,expansion=true]{microtype} % Better typography

\newcommand{\qed}{\nobreak \ifvmode \relax \else
      \ifdim\lastskip<1.5em \hskip-\lastskip
      \hskip1.5em plus0em minus0.5em \fi \nobreak
      \vrule height0.75em width0.5em depth0.25em\fi}

\usepackage{bbm}
\usepackage{MnSymbol}
\usepackage[all]{xy}

\usepackage{xcolor,colortbl,array,amssymb}

\newcommand{\x}[1]{{#1}}

\newcommand{\y}[1]{{#1}}

\makeatletter
\title{ \Large{\x{Fundamental} Nomic Vagueness}    % Title
} % Subtitle

\author{Eddy Keming Chen\thanks{Department of Philosophy,  University of California, San Diego, 9500 Gilman Dr, La Jolla, CA 92093-0119. Website: www.eddykemingchen.net. Email: eddykemingchen@ucsd.edu  }}

\date{Forthcoming in \textit{The Philosophical Review}} % Date

\begin{document}

\maketitle % Print the title section

%----------------------------------------------------------------------------------------
%	ABSTRACT AND KEYWORDS
%----------------------------------------------------------------------------------------

%\renewcommand{\abstractname}{Summary} % Uncomment to change the name of the abstract to something else

\begin{abstract}
If there are fundamental laws of nature, can they fail to be exact? In this paper, I consider the possibility that some fundamental laws are  vague. I call this phenomenon \emph{\x{fundamental} nomic vagueness}. \x{I characterize fundamental nomic vagueness as the existence of borderline lawful \emph{worlds} and the presence of several other accompanying features.} \y{Under certain assumptions, such vagueness prevents the fundamental physical theory from being completely  expressible in the mathematical language. Moreover, I suggest that such vagueness can be regarded as \textit{vagueness in the world}.}

For a case study, we turn to the Past Hypothesis, a postulate that (partially) explains the direction of time in our world. We have reasons to take it seriously as a candidate fundamental law of nature. Yet it is  vague: it admits borderline (nomologically) possible worlds.  An exact version would lead to an \emph{untraceable arbitrariness}  absent in any other fundamental laws. However, the dilemma between \x{fundamental} nomic vagueness and untraceable arbitrariness is  dissolved in a new quantum theory of time's arrow.

\end{abstract}

\hspace*{3,6mm}\textit{Keywords:  vagueness,  exactness, higher-order vagueness, mathematical expressibility, epistemicism, arbitrariness, imprecise probability, laws of nature, time's arrow, Past Hypothesis, entropy, fundamentality, Humeanism, anti-Humeanism, density matrix}   % Keywords

%\newpage

\begingroup
\singlespacing
\tableofcontents
\endgroup

%\vspace{30pt} % Some vertical space between the abstract and first section

%----------------------------------------------------------------------------------------
%	ESSAY BODY
%----------------------------------------------------------------------------------------

%Blah blah \citet{PaulNREWM} \citep{PaulNREWM}  

%------------------------------------------------
\nocite{wilson2013determinable, frege1973begriffsschrift, russell2009philosophy, wilson2013determinable, raffman1994vagueness, miller2019fundamental}

\section{Introduction}
\x{Vagueness is a pervasive feature of ordinary language.} Many predicates we use in everyday contexts do not have determinate boundaries of application.  Is John bald when he has exactly 5250 hairs on his head? There are determinate cases of ``bald,'' but there are also  borderline cases of ``bald.'' In other words,  predicates such as ``bald'' are indeterminate: there are individuals  such that it is indeterminate whether they are bald.\footnote{There are subtleties  about how best to characterize vagueness.  For  reviews on vagueness and the sorites, see  \cite{KeefeSmith1}, \cite{sep-vagueness}, and \cite{sep-sorites-paradox}. } Moreover, the boundaries between  ``bald'' and ``borderline bald'' are also indeterminate. Hence, there do not seem to be sharp boundaries \emph{anywhere}. The phenomenon of vagueness gives rise to many paradoxes (such as the sorites) and serious challenges to classical logic. 

\x{We might expect} that, at the level of fundamental physics, the kind of vagueness that ``infects'' ordinary language should disappear. That is, the fundamental laws of physics, the predicates they invoke,  and the properties they refer to should be exact.\footnote{ \x{\cite{PenroseRR} seems to hold that expectation. I used to expect that, but considerations of the arguments in this paper convinced me otherwise.} Here, we assume that there are fundamental laws of nature and it is the aspiration of physics to discover them. At the level of non-fundamental physics, and in the special sciences, the ideal of exactness may still be important but not absolute.}  The expectation is supported  by the history of physics and  the ideal that physics should deliver an objective and precise description of nature. All the paradigm cases of candidate fundamental laws of nature are not only simple and universal, but also \emph{exact}, in the sense that, for every class of worlds (or class of solutions), fundamental laws either determinately apply or determinately fail. Suppose the fundamental laws are Newton's equation of motion $F=ma$ and law of universal gravitation $F=Gm_1m_2/r^2$: there is no   vagueness about whether a certain class of worlds (described in terms of   trajectories of point particles with Newtonian masses) satisfies the conjunction.\footnote{\y{An alleged violation of the expectation concerns the measurement axioms of orthodox quantum theory, which I discuss in \S2.5. Another case concerns  the constants and the observables of effective field theory (EFT).  See \cite{miller2019fundamental}. Since he is not focused on ``fuzziness,'' Miller's proposal is best construed as a theory of ontic imprecision, not ontic  or fundamental nomic vagueness.} }

Fundamental nomic exactness---the ideal (roughly) that fundamental laws are exact---supports an important principle about the (perfect) mathematical expressibility of fundamental laws. If some fundamental laws were vague, it would be difficult to describe them mathematically in a perfectly faithful way that genuinely respects  their vagueness  and does not impose  sharp boundaries anywhere. The kind of mathematics we are used to, built from a set-theoretic foundation, does not lend itself naturally to model the genuine fuzziness of vagueness. One could go further: the language of mathematics and the language of fundamental physics are supposed to be exemplars for  the ``ideal language,'' a language that is exact, suggested in Frege's \emph{Begriffsschrift},  Russell's logical atomism, and \y{Leibniz's} \emph{characteristica universalis}. The successful application of mathematical equations in formulating physical laws \textit{seems} to leave no room  for vagueness to enter  into a fundamental physical theory. \y{If vagueness is not perfectly mathematically expressible, and if there is fundamental nomic vagueness, then the fundamental physical theory is not perfectly mathematically expressible.}\footnote{\y{This may cast some doubt on what \cite{wigner1960unreasonable}  calls the ``unreasaonable effectiveness of mathematics in the natural sciences.''}  } 

Interestingly, little is written about the connection between vagueness and fundamental laws of nature. The topic is philosophically and scientifically important, with ramifications for  metaphysics,  philosophy of science, and foundations of physics.   What does it mean for a fundamental law to be vague?  Are there examples of vague fundamental laws that may obtain in a world like ours? What does \x{fundamental} nomic vagueness mean for the metaphysical status and mathematical expressibility of fundamental laws? How does it relate to ontic vagueness? This paper is an attempt to address some of those questions. 

 First (\S2), I propose an account of \x{fundamental} nomic vagueness, distinguish it from approximations, and discuss its implications for nomic possibility and necessity. \y{Although it differs from standard cases of ontic vagueness, it can be regarded as ``vagueness in the world.'' Fundamental nomic vagueness violates a principle called \textit{fundamental exactness} on certain anti-Humeanism but not on Humeanism about laws of nature.}  (The first part of \S2.1 may be skimmed by experts \x{in}  vagueness.)

Second (\S3), I focus on the case of the Past Hypothesis (PH), the postulate that (roughly) the universe started in a special macrostate of low entropy. Given its role in explaining the arrows of time in our world, we have reasons to take it seriously as a fundamental law or at least an \emph{axiomatic postulate} in physics that is on a par with fundamental laws. Yet, macrostates are vague. Even when we specify an exact level of entropy, PH remains vague: there will be borderline lawful worlds with features of genuine fuzziness. An exact version of PH (which I call the Strong Past Hypothesis) contains an objectionable kind of arbitrariness not found in any other fundamental laws or dynamical constants---its exact boundary leaves no trace in the material ontology,  resulting in a gap between the nomic and the ontic. It violates a theoretical virtue that I call \emph{traceability}. (\S3.1  may be skimmed by experts \x{in} statistical mechanics.)

The case study highlights a dilemma between  \x{fundamental}  nomic vagueness and untraceability. In \S4, I suggest that, under some conditions, the dilemma can be dissolved in a natural way. The conditions are realized in a new quantum theory of time's arrow that links  the macrostate directly to the micro-dynamics. Surprisingly, far from making the world fuzzy or indeterminate, quantum theory can restore exactness in the nomological structure of the world. \x{(\S4 may be skipped by people whose main interests lie outside philosophy of physics.)}

\section{\x{Exactness and Vagueness of the Fundamental Laws}}

In this section, I  propose an account of  exactness and  vagueness of the fundamental laws.  
%I start from the phenomenon of vagueness in ordinary language. I propose a similar account of  (one species of) nomic vagueness and nomic exactness, in terms of the existence and non-existence of borderline worlds. I distinguish those notions from the kinds of issues that arise from approximations and the applications of fundamental laws to subsystems of the universe. I also address the connections to semanticism and epistemicism about vagueness, and the implications  for Humeanism and anti-Humeanism about laws of nature. 
%An obvious example of nomic vagueness lies in the quantum measurement axioms but it is too remote a possibility to be taken seriously. This will provide the theoretical background as we focus on a concrete and more realistic case of nomic vagueness in \S3. 

\subsection{What They Are}

First, I review some features of vagueness in ordinary language predicates. They have analogues in  \x{fundamental} nomic vagueness.  The paradigmatically vague predicates include ones such as ``bald,'' ``tall,'' ``red,'' ``child,'' and ``heap.''  Following \cite{KeefeSmith1}, I summarize their common features:
\begin{description}
	\item[(Borderline)] Vague predicates  have borderline cases. 
\end{description}
To be a borderline case is to be some object or state of affairs for which the predicates do not determinately apply. John with exactly 5250 hairs on his head is a borderline case of ``bald.'' %The parenthetical qualifier `apparently' allows  for views  according to which borderline cases do not really exist. On  epistemicism, there \emph{seems} to be borderline cases of baldness because we lack precise knowledge about where the determinate boundaries are. 

\begin{description}
	\item[(No Sharp Boundary)]  Vague predicates do not have well-defined extensions. 
 \end{description}
A precise extension of ``bald'' and a precise extension of ``not bald'' would pick out a precise boundary between the two. Suppose anyone with  6000 or more hairs is not bald and anyone with fewer than 6000 hairs is bald. Then John would fall under the extension of ``bald.'' Alex with exactly 6000 hairs would fall under the extension of  ``not bald,'' but if Alex loses just one more hair she would fall out of  ``not bald'' and fall into  ``bald.'' \x{(We can add a parenthetical qualifier `apparently' to accommodate views such as epistemicism according to which vague predicates do have well-defined extensions but we don't know what they are.)}   %????? quotation mark around apparently is  used correctly? 

% If \emph{being bald} and \emph{being non-bald} are mutually exclusive and jointly exhaustive, then given a well-defined extension of \emph{being bald} there will be a well-defined extension of \emph{being non-bald}. If they are mutually exclusive but not jointly exhaustive, there may be intermediate cases, such as \emph{being borderline bald} and \emph{being borderline non-bald}. But again, it seems completely arbitrary to draw an exact line anywhere, even between any of these ``borderline predicates.'' 

\begin{description}
	\item[(Sorites)]  Vague predicates are susceptible to sorites paradoxes. 
 \end{description}
 It is easy to generate sorites paradoxes on vague predicates. For example, we can start from a case that is determinately bald (having no hair) and proceed to add  one hair at a time,  argue that at no point can adding one hair make the difference between  ``bald'' and  ``not bald,'' and come to the absurd conclusion that no number of hairs will make one non-bald.  %?????? 

\begin{description}
	\item[(Higher-order Vagueness)]   \x{Vague predicates come with higher-order vagueness. }
 \end{description}
 Whenever there are  borderline cases, there are  borderline borderline cases, and  borderline borderline borderline cases. This is known as the phenomenon of higher-order vagueness. If it is indeterminate where to draw the line between ``bald'' and ``not bald,'' plausibly it is indeterminate where to draw the line between ``bald'' and ``borderline bald,'' and between ``not bald'' and ``borderline not bald,'' and so on. In other words, it seems inappropriate to draw a sharp line \emph{at any level}. This is part of the genuine fuzziness we are interested in below. 
 
 Higher-order vagueness is a challenge to any formal and precise model of vagueness. Even  on degree-theoretic accounts of vagueness, there will be an exact boundary between maximal determinateness and less-than-maximal determinateness and exact boundaries around  any determinate degree of vagueness, which seems unfaithful to the phenomena of higher-order vagueness \x{\cite[pp. 46-47]{KeefeSmith1}}.  The same point applies to imprecise probabilities that are treated in terms of set-valued measures. After all, a set of probability measures  is still too precise to faithfully represent the phenomenon of higher-order vagueness.\footnote{ See \cite{rinard2017imprecise} for  insightful arguments  against using set-valued probabilities to model  imprecise probabilities (IP).  Rinard's argument is relevant to our discussion of PH. Even if we use a probability distribution or a set of probability distributions concentrated on some macrostate, it is still too precise. To genuinely respect higher-order vagueness, we can replace a set of probability distribution with a vague ``collection'' of probability distributions, where some distributions will be borderline members of the ``collection.'' Membership turns out to be vague.  } \x{Even some defenders of the degree-theoretic accounts acknowledge that the numbers used to model vagueness should be taken instrumentally and not realistically. For example, \cite{edgington1996vagueness} suggests that ``[the] numbers serve a purpose as a theoretical tool, even if there is no perfect mapping between them and the phenomena'' (p.297). Can higher-order vagueness be mathematically expressed in a completely faithful way (with a perfect mapping between the mathematical representation and the phenomena)? I doubt it, but I do not have an impossibility \textit{proof} (although \cite{sainsbury1990concepts} gives compelling arguments that higher-order vagueness cannot be faithfully described set-theoretically).\footnote{\x{ \cite{WilliamsonVagueness} \S4.12 further argues that higher-order vagueness is not faithfully modeled by many-valued logics (and degree-theoretic approaches to truth).}}  I  shall assume that it cannot be. If my assumption is incorrect, then this paper can be seen as another reason to look for a perfect mathematical representation.}

We might expect that, at the fundamental level of reality, everything is perfectly exact. In particular, we might expect that there is no vagueness in the fundamental physical ontology of the world (the fundamental physical objects and their properties) or  in the fundamental nomological structure of the world (the fundamental laws).

\begin{figure}
\centerline{\includegraphics[scale=0.2]{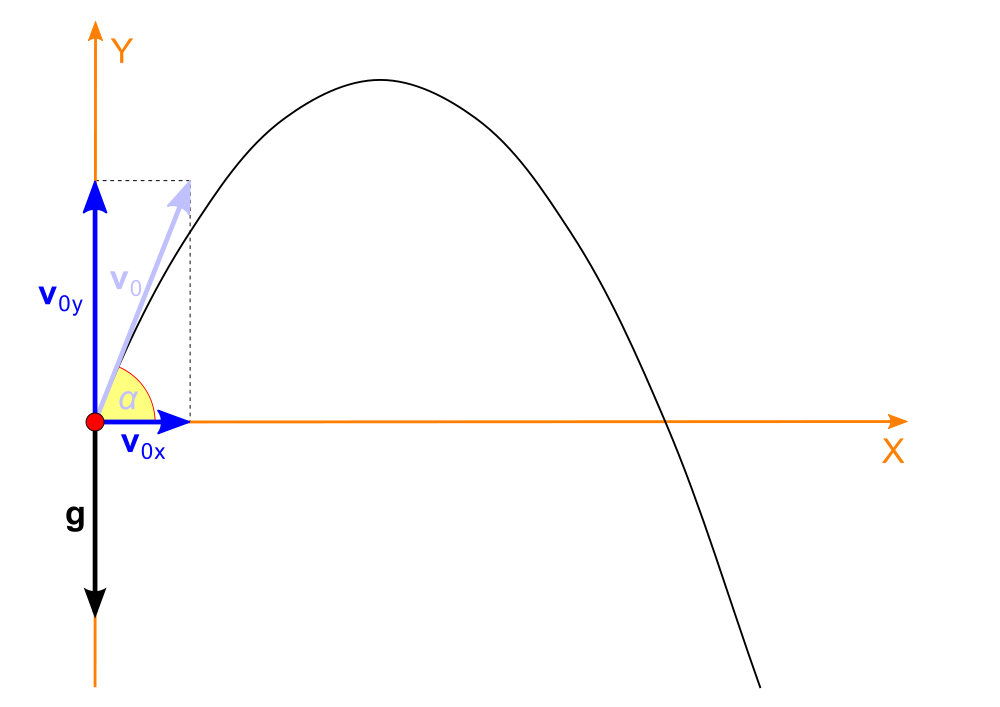}}
\caption{The motion of a projectile under Newtonian mechanics. \y{Picture by Zátonyi Sándor, (ifj.) Fizped, CC BY-SA 3.0 <https://creativecommons.org/licenses/by-sa/3.0>, via Wikimedia Commons.} }
\end{figure}

How should we understand the exactness of  paradigm fundamental laws of nature? Let us start with  the familiar case of Newtonian mechanics with Newtonian gravitation.  The theory can be formulated as a set of differential equations that admit a determinate set of solutions. Those solutions will specify all and only the possible histories compatible with  Newtonian equations ($F=ma$ and $F=Gm_1m_2/r^2$); each solution  represents a nomologically possible world  of the theory.

Consider the projectile motion illustrated in Figure 1. Suppose that the projectile has unit mass $m$ and the gravitational acceleration is  $g$ (we simplify the example by ignoring the rest of the world). We can specify the history of the projectile with the initial height, initial velocity, maximum height, and distance traveled. There is a determinate set of histories compatible with  Newtonian equations. For any history of the projectile, it  is either determinately compatible with the equations or determinately incompatible with the equations. And the same is true when we fully describe the example by accounting for all the massive bodies in the world.

 In terms of possible worlds (or models, if one dislikes possible worlds): if $W$ represents the space of all possible worlds, then Newtonian mechanics corresponds to a proper subset in $W$ that has a determinate boundary, \x{where the boundary is not in spacetime but in modal space.} Let us call that subset the \emph{domain} of Newtonian mechanics; it represents the nomological possibilities according to Newtonian mechanics. For any possible world $w \in W$, either $w$  is  contained in the domain of   Newtonian mechanics or it is not. For example, in Figure 2, $w1$ is inside but $w2$ is outside the set of worlds delineated by Newtonian mechanics. In other words, $w1$ is nomologically possible while $w2$ is nomologically impossible if Newtonian laws are true and fundamental.  This suggests that we can capture an aspect of  \x{fundamental} nomic exactness in terms of domain exactness: 

\begin{description}
	\item[Domain Exactness] \x{A fundamental law $L$ is domain-exact if and only if, (a) for any world $w \in W$, there is a determinate fact about whether $w$ is contained inside $L$'s domain of worlds, i.e. $L$'s domain has no borderline worlds, (b) $L$'s domain, which may also be called $L$'s \textit{extension}, forms a set of worlds,   (c) $L$'s domain is not susceptible to sorites paradoxes, and (d) $L$'s domain has no borderline borderline worlds, no borderline borderline borderline worlds, and so on.}  
\end{description}

\begin{figure}
\centerline{\includegraphics[scale=0.22]{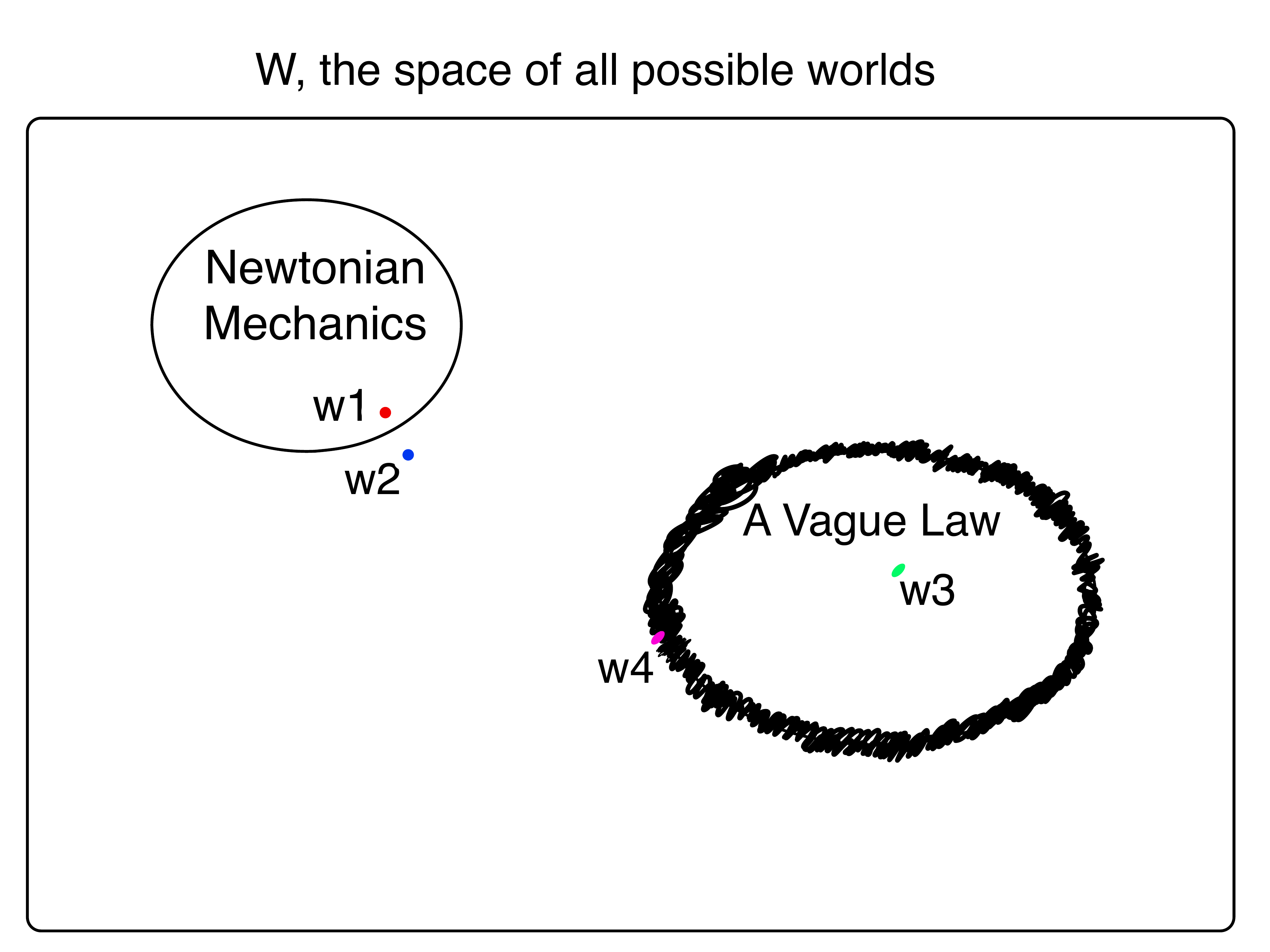}}
\caption{An exact fundamental law and a vague fundamental law represented in modal space. }
\end{figure}

\x{In contrast, a domain-vague law has none of (a)--(d). Intuitively, a domain-vague law has a vague boundary in the following sense.} In Figure 2, a domain-vague law is pictured by a ``collection'' of worlds with a fuzzy boundary. Just as a cloud does not have a clear starting point or a clear end point, the fuzzy ``collection'' of worlds does not delineate the worlds into those that are clearly compatible and those that are clearly incompatible with the law. To borrow the words of \cite{sainsbury1990concepts}, a domain-vague law classifies  worlds ``without setting boundaries'' in modal space. For example, $w3$ is clearly contained inside the domain of the vague law, since it is so far away from the fuzzy boundary; but $w4$ is not clearly contained inside the domain of the vague law, and neither is it clearly outside; $w2$ is clearly outside the domain. \x{More precisely, I propose that we understand domain vagueness as the opposite of domain exactness: }

\begin{description}
	\item[Domain Vagueness] \x{A fundamental law $L$ is domain-vague if and only if $L$ meets all four conditions below.}  
\end{description}
\begin{itemize}
	\item[ (a')] $L$ has borderline worlds that are not determinately compatible with it. For some world $w\in W$, there fails to be a determinate fact about whether $w$ is contained inside $L$'s domain of worlds. 
	\item[(b')] $L$  lacks a well-defined extension in terms of a set of models or a set of nomological possibilities. Nomological necessities and possibilities turn out to be vague. 
	\item[(c')] $L$ is susceptible to sorites paradoxes. We can start from a world that is determinately lawful, proceed to gradually make small changes to the world along some relevant dimension, and eventually arrive at a world that is determinately unlawful. But no particular small change makes the difference between lawful and unlawful. 
	\item[(d')] $L$ possesses higher-order domain-vagueness. Whenever there are borderline lawful worlds, there are borderline borderline lawful worlds, and so on. It seems inappropriate to draw a sharp line anywhere. This reflects the genuine fuzziness of domain vagueness.
	%\footnote{Again, a set-valued approach would still be too sharp to model the vagueness here. For example, for a domain-vague law $L$, one could group the possible worlds into three sets: possible, borderline possible, and impossible. Alternatively, one could assign exact numerical degree of possibility from the continuous range between 1 (maximally possible) and 0 (maximally impossible). But neither would respect the phenomenon of higher-order vagueness or the genuine fuzziness we are interested in here. }  
	%\x{It is  fine to hold an instrumentalist or anti-realist view about the mathematical representation of vagueness using exact numbers (degrees of truth, degrees of membership, degrees of possibility). For example, Edgington (CITE), a defender of the degree-theoretic approach to vagueness, writes (p.???) that ``[the] proposal does not eliminate vagueness. The numbers serve a purpose as a theoretical tool, even if there is no perfect mapping between them and the phenomena; they give us a way of representing significant and insignificant differences, and the logical structure of combination of these.''}
\end{itemize}
Thus, domain vagueness  has  features similar to those of ordinary-language vagueness. 
\x{What if $L$  has some but not all of (a')--(d')? For example, a law may have a domain where it is indeterminate whether some world $w$ is contained in it, but its domain does not have higher-order vagueness or sorites-susceptibility. If such a case exists, it may be a case of  indeterminacy, but I do not characterize it as vagueness.}\footnote{\x{ In this paper, by `a fundamental law fails to be exact,' I mean it  satisfies all of (a')--(d'). It is plausible that (a') implies (b'), (c'), and arguably (d'). However, I do not argue for that here.  } }
At any rate, domain exactness and domain vagueness capture the kind of  \x{fundamental} nomic exactness and  \x{fundamental}  nomic vagueness we care about in this paper.\footnote{There is another kind of  \x{fundamental}  nomic vagueness that results from vague objective probabilities or typicalities. See \cite{goldstein2012typicality}. \cite{fenton2019imprecise} offers an account of imprecise (but not vague) chances in the best-system theory. } We will use them to understand some case studies in the following sections. In \S3, we show that PH, if it is true and if it can be regarded as a fundamental law, is an instance of a vague fundamental law of nature. It exemplifies domain vagueness: the ``collection'' of worlds compatible with PH does not have a sharp boundary.

\subsection{What They Are Not}

%[revise in light of value vagueness.]

To better understand  \x{fundamental} nomic exactness and  \x{fundamental} nomic vagueness, it would be helpful to say what they are not. \x{Importantly, they are not about the absence or the presence of approximations. As pictured in Figure 2, the worlds that make up the domain of fundamental laws are  entire possible worlds (conceived as complete world histories, or histories of the universe).} Approximations  arise when we apply fundamental laws to partial world histories, such as histories of some subsystems of the universe. 
 In many subsystems, they are governed by effective laws---laws that  are only approximately true about certain kinds of subsystems. When a subsystem is not completely isolated from its environment (the rest of the universe), there may be forces between objects in the subsystem and objects in the environment that are negligibly small but nonzero. In that case, we can, for all practical purposes, treat the subsystem as if it were a closed system and still apply the fundamental laws to the subsystem, but with the understanding that such laws are only \textit{approximately} true. On my view, that is not  \x{fundamental} nomic vagueness.  In that case, the fundamental laws are determinately false about the subsystem.

\subsection{Laws of Nature}

\x{So far we have focused on exactness and vagueness of fundamental laws. The issue can also come up  in non-fundamental scientific theories, which can employ vague non-fundamental laws.} For example,  in so far as biology has laws, they may invoke vague predicates such as ``cell,'' ``organism,'' and ``species.''  The focus and the novelty of this paper lies in developing (in \S2.1) a general account of nomic vagueness of \textit{fundamental laws of nature} and arguing (in \S3) that PH is a realistic case  of such nomic vagueness.   Vagueness in the fundamental laws  is surprising and deserves special attention from metaphysicians and philosophers of science. If some vague fundamental laws are part of the complete theory of the physical world, and if  such vagueness is not fully mathematically expressible, then the complete theory  cannot be faithfully written in the language of mathematics. Nonetheless, the discussion may also provide a model for understanding nomic vagueness at the non-fundamental levels, which I leave for future work. 

%For the purpose of this paper, I consider laws of nature to be objective nomological structures of the world. The ontology of laws may be parasitic on  the material ontology of particles and fields (the Humean mosaic). On Humean views, laws supervene on the mosaic. Even though laws are not metaphysically fundamental on the Humean view, they are still objective.  On some anti-Humean views,   laws are distinct from the material ontology, and the former \textit{govern} the latter.  Some of the nomological structures can be represented by natural language sentences (such as English) with the help of mathematical symbols. But if there are vague laws, the representational capacity of such languages will be limited. For example, sentences expressing strict mathematical relations will draw sharp boundaries somewhere, which is unfaithful to the phenomenon of higher order (nomic) vagueness. 

Let me say more about the notion of fundamental laws employed in this paper\footnote{Similar conceptions can be found in the writings of physicists such as \cite{steven1992dreams} and \cite{hawking2008} as well as philosophers of physics \cite{albert2000time, albert2015after} and \cite{loewer2012two, loewer2016mentaculus}. This conception of laws is related to \textit{reductionism} of non-fundamental sciences to physics. On reductionism vs. autonomy of ``special sciences,'' see for example \cite{oppenheim1958unity} and \cite{FodorSS}. On my conception, even though non-fundamental laws  are derivable from fundamental laws, non-fundamental laws are not redundant; they play important roles in scientific explanations at non-fundamental levels. This conception might not contradict the viewpoint of \cite{FodorSS}. }: 

\begin{description}
  \item[Fundamental Laws] For any world $w$, fundamental laws of nature in $w$ are the non-mathematical axioms (basic postulates) of the  complete fundamental physical theory of $w$.\footnote{We also expect the complete fundamental theory to be simple, consistent, and unified.} 
\end{description}
Examples of physical theories  include classical mechanics and quantum mechanics.\footnote{ The true theory of the world is not yet discovered. But we have some hints of what it might look like, based on current theories.  } For a physical theory to be complete in world $w$, it needs to entail all the important regularities in $w$, including those described by  non-fundamental laws (such as laws of chemistry, biology, and so on). For a physical theory to be fundamental in $w$, it cannot be derived from another \x{non-equivalent} physical theory that is true in $w$.\footnote{\x{The qualifier ``non-equivalent'' is added so that merely having an equivalent reformulation does not disqualify a theory from being fundamental. Without the qualification, if $T$ and $T'$ are two equivalent physical theories that are mutually derivable, then neither $T$ nor $T'$ can be fundamental. Thanks to an anonymous reviewer for suggesting I make this clear.}} Hence, in a quantum world, classical mechanics is not a fundamental theory, because it can be  derived  from quantum mechanics (via approximations in some limit). However, in a classical world, classical mechanics is a fundamental theory  because quantum mechanics is not true in such a world. In \S3.1, I suggest that in a time-asymmetric classical world, classical mechanics by itself is not \textit{complete} because it fails to account for the time asymmetries, which is an important class of regularities (and the same applies to a time-asymmetric quantum world). A solution is to add PH to the micro-dynamical equations to complete the fundamental theory.  Fundamental laws have an elite status---they are  axiomatic. In terms of modality, since the axioms are nomologically necessary, the derived theorems will also have nomological necessity. The derivation may be complicated due to  mathematical complexities.
The physical theory can employ some mathematics, but the mathematical axioms are not laws of nature. Hence, the fundamental laws of nature are the \textit{non-mathematical axioms}. 

Laws in $w$ are either fundamental  or non-fundamental in $w$. I require  non-fundamental laws in $w$ be derivable\footnote{The relevant derivations may also involve approximations and idealizations. Again, we set these issues aside as it does not impact the argument in \S3. } from  fundamental laws in $w$. But not all deductive consequences of fundamental laws are laws, for otherwise we could  trivialize the notion of laws by  using disjunction introduction.  Some deductive consequences  will be more important than others because they support counterfactuals and are extraordinarily useful and simple. Identifying the sufficient conditions for laws is an important project,  but I do not pursue it here. Instead, I suggest a necessary condition for a law to be non-fundamental in $w$: 
\begin{description} 
  \item[Necessary Condition for Non-Fundamental Laws] In any world $w$, if a law of nature is a non-fundamental law in $w$, then it can be (non-trivially) derived from the fundamental laws in $w$. 
\end{description}
Consequently, a law that cannot be so (non-trivially) derived in $w$ is a fundamental law in $w$. An example of a non-fundamental law in our world is the ideal gas law $PV=nRT$ that can be derived from the micro-physics.  Not all non-fundamental laws have been successfully derived from the fundamental axioms in physics, but what matters is that they can be.  %I use the above conception of laws of nature  in \S3 to show that PH is a fundamental law as it is a law of nature and cannot be derived from other fundamental laws. 

The above conception is somewhat neutral about the metaphysics of lawhood. Let me say more about the metaphysical commitments. With \cite{hicksschaffer}, I do not insist on \cite{LewisNWTU}'s strict criterion that fundamental laws  are stated exclusively with predicates that correspond to perfectly natural properties.\footnote{See \cite{SiderWBW} for a defense and  expansion of Lewis's criterion. }  When formulating fundamental laws, scientists should be free to invoke derived properties  as long as they are scientifically useful.  In fact,  Lewis's own  analysis of probabilistic laws allows \textit{objective probability} to appear in the axioms of the best system yet objective probability is not perfectly natural (and neither is it categorical).\footnote{Thanks to Chris Dorst for discussion here.}
For the Humean best-system framework,  I suggest we replace the strict criterion  with this one:  the best system optimally balances simplicity, informativeness, fit, and \textit{degree of naturalness} of the properties invoked by the axioms.\footnote{For similar suggestions see \cite{hicksschaffer, fenton2019imprecise}, and \cite{sanchezcrystallized}. It would certainly be useful to give a full account on how to calculate the degree of naturalness, but I do not pursue it here. }  The revision   arguably still handles the trivialization problem  that motivates the strict criterion.\footnote{ \x{Briefly, the trivialization problem is this \cite[p.367]{LewisNWTU}: without some constraint of naturalness, the best system can be just the simple and maximally informative sentence $\forall x Fx$ where  $F$  applies to all and only actual objects. The system entails all actual truths and regularities. That result, on Lewis's  account, means all regularities are laws, which trivializes the  distinction between laws and non-laws. Lewis appeals to perfect naturalness to block the problem, but  degree of naturalness also suffices \citep{hicksschaffer}.}} In so far as derived properties are instantiated by fundamental objects, there is nothing metaphysically spooky. For example, even though acceleration is a derived property (from positions) in Newtonian mechanics, $F=ma$ can  delineate what is nomologically possible. Similarly, entropy is a derived property (from volume in phase space), and a low-entropy initial condition can also constrain the nomological possibilities. As such, fundamental laws involving derived properties can still  describe (or govern) the behavior of fundamental objects. Nevertheless, in the cases of entropy and  acceleration, unlike the property of \textit{being a tiger}, the derived properties are easily re-expressible in terms of fundamental properties and the re-expressed laws can still be simple. Hence,  sometimes even the strict criterion can be satisfied (see \S3.2).

Moreover,  with \cite{callender2004measures} and \textit{pace} \cite{LewisNWTU}, I do not require laws to concern exclusively  \textit{regularities}. Certain boundary conditions of the universe, such as PH, are candidate axioms of the Lewisian best system. On Humeanism, why automatically disqualify PH, which seems to  be an axiom of the actual best system of our world, from being a fundamental law?\footnote{Some may worry that this dissolves the distinction between boundary conditions and laws. However, not all boundary conditions have the required balance of optimally contributing to simplicity, informativeness, fit, and degree of naturalness.} I see no compelling reason. %After all, even \cite[p.368]{LewisNWTU} regards fundamental laws as the axioms of the best system.  
There are many other versions of Humeanism, such as those found in \cite{cohen2009better} and \cite{loewer2020package}. For concreteness, when I discuss Humeanism, I  focus on this particularly interesting version of the  non-governing conception of laws:
\begin{description}
  \item[Reformed Humeanism] The fundamental laws are the (non-mathematical) axioms of the best system that summarizes the mosaic and optimally balances simplicity, informativeness, fit, and degree of naturalness of the properties invoked. The best system supervenes on the mosaic.\footnote{Here I set aside the issue whether the Humean mosaic contains only categorical  and ``point-sized'' facts.} 
\end{description}

Regarding the governing conception of laws, with \cite{MaudlinMWP}, I think the best versions  are ones where laws are interpreted as ontological primitives; they are not to be analyzed further in terms of universals, essences, powers, dispositions, and the like.  However, \textit{pace} Maudlin, I do not posit the direction of time as a fundamental feature of the universe and do not consider temporal \textit{production} as central to the notion of governing. Consequently, for anti-Humeanism, I would not restrict fundamental laws to dynamical laws of temporal evolution. In particular, boundary conditions can be fundamental  laws. A boundary condition such as PH can govern the world by \textit{constraining} the nomological possibilities, thereby limiting the range of behavior of fundamental objects. Hence, I am particularly interested in a minimal governing conception of laws:

\begin{description}
  \item[Minimal Primitivism] Fundamental laws of nature are the (non-mathematical) fundamental facts that govern the behavior of fundamental objects; there is no restriction on the forms of the fundamental laws.   
\end{description}
The theoretical virtues invoked by the reformed Humean are still useful for the minimal primitivist: 
\begin{description}
  \item[Epistemic Guides] Even though theoretical virtues such as simplicity, informativeness, fit, and degree of naturalness are not metaphysically constitutive of fundamental laws, they are good epistemic guides for discovering them. 
\end{description}
Reformed Humeanism and minimal primitivism are two well-motivated metaphysical views about laws.  \x{However, they are not accepted by everyone.} I do not have the space to defend them in detail, but I think they capture two important metaphysical conceptions of lawhood  in the literature.  
 In \S3.1, I show that both support the proposal that PH is a fundamental law.

\subsection{Vagueness in the World?}

\x{Fundamental laws of nature are objective features of the physical world. Thus,  \x{fundamental} nomic vagueness  appears to be  ``worldly.'' However,  \x{fundamental} nomic vagueness differs from  standard cases of worldly or \textit{ontic} vagueness that concern the vague identity, spatio-temporal boundaries, and parts of material objects (such as cats, clouds, mountains, and tables).\footnote{\x{For an overview of ontic vagueness, see \cite{sep-vagueness} \S8, \cite{KeefeSmith1, williams2008ontic}, and the references therein.} } That is because fundamental laws of nature are not material objects and do not have boundaries or parts in spacetime. Moreover,  fundamental nomic vagueness is not modeled by Barnes's (2010) theory,  one of the most developed theories of ontic vagueness to date. Nevertheless, if fundamental laws are fundamental facts of the world, then their vagueness \textit{can} be seen as metaphysical. Hence, there is conceptual room for a more liberal conception of ontic vagueness that includes fundamental nomic vagueness. 

Standard cases of ontic vagueness  are  material objects with vague boundaries or parts in spacetime.\footnote{\x{See \cite{tye1990vague}, as well as related discussions about `the problem of the many' in \cite{GeachRG} and \cite{UngerPM}. See \cite{HawleyHTP} for a discussion about vagueness in temporal persistence. }} For concreteness, consider Tom the cat. Tom is composed of many molecules, but some  are borderline parts of Tom (and some are borderline borderline parts of Tom, and so on). Moreover, there are many exact groupings of the molecules, called p-cats. Which p-cat is identical to Tom? It may be vague. We can also run sorites argument and remove molecules one at a time from Tom. As a first approximation, the picture suggested by the focus on standard cases seems to be this\footnote{\x{Some of the theorists such as \cite{parsons1995worldly} and \cite{barnes2010ontic} mention only   indeterminacy in their official statements even though their target is ontic vagueness. It does not matter to my arguments below. If such indeterminacy does not come with other features of the target sense of vagueness (sorites-susceptibilty, higher-order vagueness, etc),  then fundamental nomic vagueness is not modeled by their theories.}}: 

\begin{description}
  \item[V1] There is ontic vagueness if and only if there is some material object (or objects) and some property (or relation) of material objects such that it is vague whether the object (or objects) has the property (or relation).\footnote{ \x{\cite{parsons1995worldly} seem to hold such a view, even though their emphasis is on the  states of affairs about   material objects having  properties or relations. For them, the objects  are those things that make up the world, and the relations include the identity relation, which is the focus of their analysis.} } 
\end{description}
On V1, if Tom is an actual vague material object, then there is ontic vagueness.  In contrast, \cite{KeefeSmith1} suggest that vagueness of macroscopic objects such as Tom is  ``merely superficial'' (p.56). Facts about Tom  supervene on the material objects and their properties at the ``base level'' (the metaphysically fundamental level), if such a level exists. Only vagueness at the fundamental level qualifies for ``non-superficial'' ontic vagueness.  They seem to hold onto the following view (although they do not mention relations): 
\begin{description}
  \item[V2] There is ontic vagueness if and only if there is some fundamental material object (or objects) and some fundamental property (or relation) such that it is vague whether the object (or objects) has the property (or relation).
\end{description}
V2 only looks at the fundamental  material objects (such as fundamental particles) and their fundamental properties and relations (such as mass, charge, and spatio-temporal relations).  \cite{burgess1990vague} challenges V2 and defends the idea in V1 that the existence of vague macroscopic objects  also  qualifies for ontic vagueness, because the idea allows one to ``say familiar things about the nature of concrete physical reality'' (p.286) and to respect the task of  metaphysics to ``describe each layer [of the multi-layered reality] and the relationships between them'' (p.283). However, neither  V1 nor V2 recognizes fundamental nomic vagueness as a version of ontic vagueness. 

To model ontic vagueness, \cite{barnes2010ontic} proposes the following: 

\begin{description}
  \item[V3] There is ontic vagueness if and only if every possible world is exact but it is vague which world is the actualized world.\footnote{\x{This relies on an ersatz modal theory where there is exactly one actual world,  possible worlds  are ersatz objects,  and one of them is actualized. See \cite{barnes2010ontic} pp.613-621.  } } 
\end{description}
Barnes's proposal is motivated by a desire to preserve classical logical principles such as bivalence and excluded middle. It is one of the most developed theories of ontic vagueness to date. Again,  fundamental nomic vagueness does not imply ontic vagueness in the sense of V3, as a fundamental law (such as  PH) may be vague without it being vague which possible world is actualized. What fundamental nomic vagueness violates is this principle: 

\begin{description}
  \item[V4] It is vague which worlds are nomologically possible.\footnote{\x{There is an interesting similarity between fundamental nomic vagueness and standard cases of ontic vagueness of material objects. While Tom has a fuzzy boundary in spacetime, a vague fundamental law (see Figure 2) has a fuzzy boundary in modal space.}} 
\end{description}
 Even so, a vague fundamental law does not contradict this: 
\begin{description}
  \item[V5] It is vague which worlds are metaphysically possible. 
\end{description} 
On a more liberal conception of ontic vagueness, according to which validating V4 suffices for there being ontic vagueness,  \x{fundamental}  nomic vagueness would then count as ontic vagueness.  On minimal primitivism, and on any view where  fundamental laws and facts about  nomologically possibilities are among the fundamental facts, validating V4 entails violating this: 
\begin{description}
	\item[Fundamental Exactness] All  the fundamental facts of the world are exact.  
\end{description}
If the idea about ontic vagueness is simply vagueness in the fundamental facts, then it seems that we should adopt the more liberal conception that leaves room for  fundamental nomic vagueness to count as ontic vagueness. However, if the idea  is more restrictive along the lines of V1--V3, then fundamental nomic vagueness does not count. }

% In fact, on a strict conception,  \x{fundamental} nomic vagueness is not ontic vagueness at all; it does not imply vagueness in the fundamental material ontology. It counts as ontic vagueness only on more liberal conceptions.  %???? problem of the many single or double quotation marks? 

%On a strict conception of ontic vagueness, what counts as ontic vagueness is vagueness in the fundamental material ontology. Assuming physicalism, the material ontology of each world (actual or merely possible) consists in the fundamental material objects  and perhaps their properties (and perhaps also their relations): particle $a$, its mass, charge, location; particle $b$, its mass, charge, location; and so on. Unlike particles and fields (which are matter), fundamental laws are not matter that make up tables and chairs; they are not part of the fundamental material ontology in any world, if they are in the fundamental ontology at all. On reformed Humeanism, fundamental laws are mere summaries of and thus metaphysically dependent on  facts about the fundamental material ontology---the mosaic. The strict conception may appeal to reformed Humeans because, for them, only the mosaic is fundamental. It may also appeal to some minimal primitivists, as the fundamental material ontology serves as the reduction base of  macroscopic objects. 

Let me say more about the connection between fundamental nomic vagueness and Fundamental Exactness. It is natural to wonder whether   \x{fundamental}  nomic vagueness would favor reformed Humeanism over minimal primitivism, since the Humean mosaic can be perfectly exact even when some  fundamental laws are vague. It depends on whether it is a vice of a theory to lose Fundamental Exactness  when it does not entail ontic vagueness in the more restrictive sense. I suspect the answer is not straightforward, as there are other factors to consider. In \S3.3, I suggest that  \x{fundamental} nomic vagueness is a relevant desideratum in theory-choice but it should be used carefully when other things are equal.   %?????

On minimal primitivism, if the epistemic guides point us to a theory with a vague fundamental axiom, then we  have good reasons to accept some  \x{fundamental} nomic vagueness. However, since the epistemic guides are defeasible and fallible, they do not guarantee finding the true fundamental laws. Some might take this as a reason to adopt an epistemic view of  \x{fundamental} nomic vagueness, according to which there is an exact boundary of nomological possibilities but it is hidden from us. However, the minimal primitivist should resist that urge, as she  has good reasons to trust the epistemic guides as truth-conducive, assuming the world is induction-friendly. Hence, if she has good reasons to take the vague postulate as a fundamental law, then  she should take that as evidence that there are indeterminate facts about nomological possibilities even if that violates Fundamental Exactness. From an empiricist perspective, metaphysical assumptions are revisable in light of empirical evidence.  

On reformed Humeanism, if the best system includes a vague axiom, then we have to accept some  \x{fundamental}  nomic vagueness. After all, on reformed Humeanism, being part of the best axiomatization is constitutive of being a fundamental law. However,  \x{fundamental} nomic vagueness does not entail a violation of fundamental exactness, as fundamental laws are not among the fundamental facts of reformed Humeanism. Even if the reformed Humean has to accept some  \x{fundamental}  nomic vagueness, she may have multiple ways to understand its nature. I leave that to future work.

\subsection{Vagueness in the Quantum Measurement Axioms?}

No physical theory has inspired more discussions about indeterminacy than quantum theory.  It has been argued that ontic vagueness in a strict sense is a feature implied by quantum theory (see, for example, \cite{lowe1994vague}, \cite{french2003quantum}). However, we now have  precise formulations of quantum mechanics  such as Bohm's theory, GRW spontaneous collapse theory, and Everett's theory (see \cite{sep-qt-issues} for a review).  In those  theories, there is no vagueness in the fundamental material ontology or fundamental dynamics (satisfying V1--V4). The world can be described as  a universal quantum state evolving deterministically (or stochastically) and  in some cases guiding and determining the trajectories of material objects, all of which are exact.

However, textbook versions of quantum mechanics seems to offer a genuine case of  \x{fundamental} nomic vagueness. We focus on the dynamical laws.\footnote{For the related issue of ``quantum metaphysical indeterminacy'' that arises from considerations of realism about quantum observables, see \cite{calosi2019quantum} and the references therein.}  Textbook versions suggest that quantum theory contains two kinds of laws: one linear, smooth, and deterministic evolution of the wave function (the Schr\"odinger equation), and the other stochastic jump of the wave function triggered by measurements of some system (collapse postulates). However, it is unclear what counts as a measurement, and hence it is unclear when the two dynamical laws apply. For any precise definition we propose, say in terms of the size of the system, we can imagine a slightly smaller or a slightly larger size that could also work. For any precise boundary between the measured system and the measuring apparatus, we can imagine a somewhat different line. So there seems to be no principled way to draw the boundary between the system and the apparatus and hence no principled definition of when to apply the two laws. 

\cite{bell1990against} speaks out against such vagueness in the fundamental axioms of quantum mechanics:
\begin{quotation}
%	It would seem that the theory is exclusively concerned about `results of measurement', and has nothing to say about anything else. 
	What exactly qualifies some physical systems to play the role of `measurer'? Was the wavefunction of the world waiting to jump for thousands of millions of years until a single-celled living creature appeared? Or did it have to wait a little longer, for some better qualified system...with a Ph.D.?....The first charge against 	`measurement', in the fundamental axioms of quantum mechanics, is that it anchors there the shifty split of the world into `system' and `apparatus'. (p.34)
\end{quotation}
Bell's first objection is that textbook quantum mechanics is too vague. Even supposing terms such as `measurement'  have determinate cases, it is hard to imagine there  be a sharp split between systems that are measurers and systems that are measured. Hence, there will be histories of the wave function that count as borderline possible. However, the real issue that troubles Bell is the disunity  suggested by the theory: the world is (vaguely) split into two parts, one classical and one quantum.  If we are allowed to apply the theory  to only part of the world, then it is ``to betray the great enterprise'' (p.34) of understanding the world in a unified way. The ``shifty split'' shows that the division is not a principled one. So it seems to me, in this case,  \x{fundamental} nomic vagueness is a \textit{symptom} that points us to the deeper problem that the theory is disunified. 
I return to this issue in \S3.3.

As a historical matter, the vagueness issue has been resolved in  precise formulations of quantum mechanics of Bohm, GRW, and Everett, making quantum mechanics less relevant as a serious case for ontic vagueness or  \x{fundamental} nomic vagueness. These theories not only resolve the vagueness issue, but they are also better physical theories. They provide deeper, more unified, and observer-independent explanations about `quantum measurements.' On those theories, measurement is  not  a \textit{sui generis} process that has special powers in the physical world, and collapse is not a process that occurs only when an observer is present. Rather, they are treated as any other process that obeys the same set of physical laws. They uphold the physicalist aspiration that observers  are just part of nature and nothing special. Hence, even setting aside the issue of vagueness, there are reasons  not to take textbook quantum axioms  as  candidate fundamental laws.  
 In order to find a more realistic case of  \x{fundamental} nomic vagueness, we must look elsewhere. 
%Surprisingly, we can already find it in the classical statistical mechanical framework of Boltzmann, and such nomic vagueness persist into  quantum statistical mechanics under the standard formulation. 

% speaks out against the orthodox quantum theory in which one of the fundamental dynamical laws about wave function collapses crucially refer to observations and measurements, which are vague terms. Certain processes may determinately count as measurements, and certain processes may determinately count as non-measurements. But due to the vagueness of the term, it is not clear where to draw the line between measurements and non-measurements.  

%The ``split'' between the measurer and the measured system is ``shifty'' and indeterminate. One way to interpret the orthodox view is that there are good ways of drawing the boundary and bad ways of drawing the boundary. For practical purposes, physicists will agree on what the measured system is, even though the precise boundary is not to be found. Indeed, it would be surprising if there were a precise boundary between the microscopic and the macroscopic. Hence, the law that employs such a notion of measurement could inherit the kind of vagueness in the ``shifty split.''  That is the case for the measurement axioms of quantum mechanics. If the boundary is vague, then there will be some worlds, with some histories of collapses, that are judged to be borderline. If they are taken to be among the fundamental laws of nature, we have a case of nomic vagueness in the sense of domain vagueness. 

\section{A Case Study of  \x{Fundamental} Nomic Vagueness: The Past Hypothesis}

In this section, I  provide a more realistic case of  \x{fundamental} nomic vagueness that arises from considerations of the arrow of time. 

%It arises when we consider the origin of time's arrow in the universe. It has been suggested that one way to explain the various asymmetries in time and to explain the Second Law of Thermodynamics is to postulate a macroscopic initial condition for the early universe. \cite{albert2000time} calls such a condition the Past Hypothesis. Given its role in the explanation of lawful regularities and its relative simplicity, it has been suggested that it is a fundamental law of nature. In \S3.1, we review the problem of time asymmetry and  explain the reasons for interpreting the Past Hypothesis as a fundamental law.  In \S3.2, we argue that the Weak Past Hypothesis---the version of the Past Hypothesis with exact values of the macroscopic variables---is vague.  In \S3.3, we show that there are reasons to regard the Strong Past Hypothesis---a non-vague version of the Past Hypothesis---as implausible.   It postulates a kind of arbitrariness that is not only unsupported by the framework of Boltzmannian statistical mechanics but also violates a plausible theoretical principle. Its arbitrariness is untraceable, unlike that other fundamental laws of nature or the physical constants they contain. In \S3.4, we explain the disanalogies with one-parameter chance hypotheses. 

\subsection{Temporal Asymmetries and the Past Hypothesis}

In a world governed by (essentially) time-symmetric dynamical laws such as  classical mechanical equations,  quantum mechanical equations, or  relativistic equations, it is plausible to think that the time-asymmetric nomological regularities (such as the tendency for entropy to increase and not decrease) cannot be derived from  dynamical laws alone. 

What else should be added? An influential proposal suggests we postulate a special initial condition: the universe was initially in a low-entropy macrostate,  one with a  high degree of order. This is now called  the \textit{Past Hypothesis }(PH).\footnote{PH was originally proposed in \cite{boltzmann2012lectures}[1898]\S89 (though Boltzmann ultimately seems to favor what may be called the \textit{Fluctuation Hypothesis}) and  discussed in \cite{feynman2017character}[1965]\S5. A geometric version was proposed by \cite{penrose1979singularities}. For recent discussions, see \cite{albert2000time}, \cite{loewer2016mentaculus}, \cite{sep-time-thermo},  \cite{north2011time}, \cite{lebowitz2008time},  \cite{goldstein2001boltzmann}, and \cite{goldstein2019gibbs}. For critical discussions, see \cite{earman2006past}. } Assuming PH and an accompanying Statistical Postulate (SP) of a uniform probability distribution over possible microstates compatible with the low-entropy macrostate, most likely the universe's entropy increases towards the future and decreases towards the past.  The presence of PH, and the absence of a corresponding low-entropy hypothesis at the other end of time, explains the wide-spread temporal asymmetry. There are several versions of PH, of varying strengths, which we discuss  in \S3.2 and \S3.3. What is important for my purpose in this section is that PH narrows down the choices of the initial microstate of the universe (a maximally fine-grained description of its physical state): they have to be compatible with some special macrostates  (coarse-grained descriptions of the physical state) with low entropy.

To qualify as a serious candidate for  \x{fundamental} nomic vagueness, PH needs to be considered as a  candidate for a fundamental law of nature.  Its nomic status has been taken seriously in the literature. For example,  \cite{feynman2017character}[1965] writes: 

\begin{quotation}
	Therefore I think it is necessary to add to the physical laws the hypothesis that in the past the universe was more ordered, in the technical sense, than it is today---I think this is the additional statement that is needed to make sense, and to make an understanding of the irreversibility.
\end{quotation}
Making a similar point, \cite{goldstein2019gibbs} write:  
\begin{quotation}
	The past hypothesis is the one crucial assumption we make in addition to the dynamical laws of classical mechanics. The past hypothesis may well have the status of a law of physics---not a dynamical law but a law selecting a set of admissible histories among the solutions of the dynamical laws.
\end{quotation}
See also  \cite{albert2000time},  \cite{callender2004measures}, and \cite{LoewerCatSLaw}.  

I think we have good reasons, both scientific and philosophical, to accept that (1) PH is a law, and (2) PH is a \emph{fundamental} law. \x{Here is another place where there is room for reasonable disagreements. Although I find the reasons compelling, others might not. They might resist (1) and (2) by rejecting my assumptions. Ultimately, my main conclusions are conditional: (A) if one accepts my conception of laws of nature, then PH is a fundamental law, and (B) if PH is a fundamental law, then there is fundamental nomic vagueness.  If someone wants to avoid committing to fundamental nomic vagueness, they might see the arguments in this paper as reasons to resist the idea that PH is a fundamental law or the conception of laws that motivates the idea.}

\x{I argue for (B) in \S3.2-3.3. For the remainder of this subsection, I argue for (A). }
I review three reasons in favor of (1) that PH is a law. 
First, the Second Law of Thermodynamics is a non-fundamental law that is derivable from PH and dynamical laws.
The Second Law is  a \textit{statistical} law: typically, most isolated subsystems increase in entropy towards the future and decrease in entropy towards the past. It can admit exceptions, which are nomologically possible but unlikely (assuming SP).  The Second Law is a statistical \textit{law} because it satisfies many desiderata for lawhood. For example, it is  projectable and supports a wide range of counterfactuals. (Even if we were on Mars, most isolated ice cubes in the cup would most likely be larger in the past and smaller in the future. Even if there were Martians, they would \x{most likely} not be able to regularly separate milk and coffee with a few casual swirls of a spoon \citep{carroll2010eternity}.)  Hence, the Second Law has nomological necessity. Now, if the PH is nomologically contingent, the Second Law will also be nomologically contingent (since the Second Law is partly derived from PH). Hence, PH is nomologically necessary. 
Second, the counterfactual arrow of time depends on nomic facts. Treating PH as a law of nature provides a good explanation for the counterfactual arrow: PH severely constrains the nomic possibility space such that typical (in the sense made precise by SP) histories exemplify counterfactual dependence in one temporal direction only. \cite{LewisCDTA}'s  explanation for the counterfactual arrow is not sufficient as shown by \cite{elga2001statistical}. As \cite{LoewerCatSLaw} points out, to fix it we  need to include PH as a law.
Third, there is an abundance of physical records about the past but no such records about the future. Having PH as a law  explains that. It also explains why such an asymmetry is not accidental or extremely unlikely. Assuming that PH is a law and SP underwrites objective probabilities, it is indeed overwhelmingly likely that our past was in a lower entropy state and not a higher entropy state like our future, thus blocking the reversibility paradox and the skeptical catastrophe (\cite{albert2000time} \S4). 

Given the conception of fundamental laws discussed in \S2.3, there are good reasons in favor of (2) that PH is a fundamental law. A law is a fundamental law in $w$ if it is not derivable from other fundamental laws in $w$. PH is not derivable from  standard fundamental  laws we postulate in our world, because they are compatible with both PH and its negation. For example, just like Newtonian mechanics, quantum mechanics is compatible with a low-entropy initial macrostate  and with a high-entropy one.\footnote{What about deriving PH from some non-standard theories? There is  progress on this front, but success is not guaranteed. For example,  \cite{carroll2004spontaneous} developed a multiverse model  with unbounded entropy in which ``baby universes'' are spontaneously created in low-entropy states. In such a model, PH is not a fundamental postulate but a local initial condition induced by  time-symmetric dynamical laws that are more fundamental. However, although theoretically possible, it is far from clear whether the actual dynamical laws produce such ``baby universes'' and produce them in sufficiently low-entropy states.
}  Hence, if PH is a law in our world, and in so far as we have good reasons to think that the true physical theory has features similar to current ones, then we have good reasons to think that PH is a fundamental law in our world. 
That, I maintain, is a reasonable verdict. After all, if PH is not derivable from other fundamental laws, it will be axiomatic in the fundamental physical theory, on a par with other fundamental laws. If PH is a non-fundamental law, how can it have an axiomatic status? If it plays the same role as other fundamental laws as axiomatic constraints on the nomological possibilities, has PH not earned the elite status?  Moreover, if one is a reformed Humean about laws of nature,  one has a direct argument for the fundamental nomological status of PH. The system containing PH and SP as axioms is way more informative  than the system without them and is only slightly more complex  (cf: \cite{callender2004measures} and \cite{loewer2012two}). Similar considerations hold for minimal primitivism, where the best system is used as our best epistemic guide to find the fundamental ``governing'' laws. The set of fundamental axioms including PH, SP, and fundamental dynamical laws has been named  \textit{the Mentaculus} by   \cite{LoewerCatSLaw} and \cite{albert2015after}. See \cite{demarest2019mentaculus} and \cite{chen2018HU, chen2020harvard} for further developments of the Mentaculus. 

One could raise two objections on metaphysical grounds.  First, PH concerns the boundary condition of the universe and is not a dynamical law.  As such, it is different from other candidate fundamental laws such as the Schr\"odinger equation. However, on my view, that is not a relevant difference. As discussed in \S2.3, being non-dynamical does not automatically disqualify PH from being a fundamental law. 
Second, PH concerns metaphysically non-elite properties such as entropy which should not appear in a fundamental law. However, from the perspective of reformed Humeanism and minimal primitivism, it is not absolute eliteness that matters but relative eliteness. As long as entropy is relatively elite (relatively natural and fundamental), it can appear in the reformed Humean best system and the minimal primitivist's fundamental laws.  Moreover, entropy can be fairly easily re-expressed in terms of fundamental properties. I discuss that in \S3.2.

These considerations provide good  reasons to think that, \x{conditionalized on the assumptions, we should interpret  PH as a fundamental law in our world.} 
%It is also acceptable to some ``minimal'' anti-Humeans. If the central anti-Humean intuition is just that laws don't supervene on the mosaic, why think only dynamical laws can be laws? A ``minimal'' anti-Humean would be perfectly happy to accept, on scientific grounds, a fundamental boundary-condition law such as the PH (or any boundary condition that earns its status as a law).  
If one is not willing to call PH a fundamental \emph{law}, one may still accept that, given its nomological necessity and underivability from other fundamental laws, PH  enjoys an axiomatic status in the fundamental physical theory. As such, its vagueness has the same ramifications for  nomic modalities and the mathematical expressibility of the fundamental theory.
%If PH is a candidate fundamental law, then the Statistical Postulate has the same claim to being fundamental law-like and the probabilities it prescribes has the status as fundamental objective chances. 

\subsection{Vagueness of the Weak Past Hypothesis}

Hence, we have good reasons to think that there is  \x{fundamental} nomic vagueness in our world \textit{if PH is vague}. Is PH vague? To begin, let us consider the following version of PH that is sometimes proposed:

\begin{description}
	\item[Super Weak Past Hypothesis (SWPH)] At one temporal boundary of space-time, the universe has very low entropy. 
\end{description}
SWPH is obviously vague. How low is low? The collection of worlds with ``low-entropy'' initial conditions has fuzzy boundaries in the space of possible worlds. Hence, if SWPH were a fundamental law, then we would have nomic (domain) vagueness.  

However, SWPH may not be  detailed enough to explain all the temporal asymmetries. For example, in order to explain the temporal asymmetries of records, intervention, and knowledge, \cite{albert2000time} and \cite{loewer2016mentaculus} suggest that we need a more specific condition that narrows down the initial microstates to a particular macrostate. One way to specify the macrostate invokes exact numeral values for the macroscopic variables of the early universe. Let $S_0, T_0, V_0, D_0$ represent the exact values (or exact distributions) of (low) entropy, (high) temperature, (small) volume, and (roughly uniform) density distribution of the initial state. Consider the following version of PH:  
\begin{description}
	\item[Weak Past Hypothesis (WPH)] At one temporal boundary of space-time, the universe is in a particular macrostate $M_0$, specified by the macroscopic variables $S_0, T_0, V_0$, and $D_0$. 
\end{description}
WPH is a stronger version of PH than SWPH. By picking out a particular (low-entropy) macrostate $M_0$ from many macrostates,  WPH more severely constrains the initial state of the universe. WPH is also more precise than SWPH. (Some may  even complain that the WPH is too strong and too precise.) Unfortunately, WPH is still vague. The collection of worlds compatible with WPH  has fuzzy boundaries. If WPH were a fundamental law, then we would still have nomic (domain) vagueness: there are some worlds whose initial conditions are borderline cases of being in the macrostate $M_0$, specified by the macroscopic variables $S_0, T_0, V_0$, and $D_0$.

The vagueness of WPH is  revealed when we  connect the macroscopic variables to the microscopic ones. Which set of microstates realizes the macrostate $M_0$? There is hardly any sharp boundary between those that do and those that do not realize the macrostate. A macrostate, after all, is a coarse-grained description of the physical state. As with many cases of coarse-graining, there  can be borderline cases. (To connect to our discussion in \S2.1, the vagueness of macrostates is similar to the vagueness of ``is bald'' and ``is a table.'') In fact, a macrostate can be vague even when it is specified with precise values of the macro-variables. This point should be familiar to those working in the foundations of statistical mechanics.\footnote{Commenting on the vagueness of the macrostate boundaries, \cite{LoewerCatSLaw} writes, ``Obviously, the notion of \emph{macro state} is vague and there are many precisifications that would serve the purposes of statistical mechanics.''  \cite{goldstein2019gibbs} write, ``there is some arbitrariness in where exactly to `draw the boundaries.'''} However, it is worth spelling out the reasons to understand where and why such vagueness exists.

Let us begin by considering the case of temperature, a macroscopic variable in thermodynamics. Take, for example, the macrostate of having temperature $T=273.15$K (i.e. 0\textdegree{C} or 32F). It is sometimes suggested  \emph{without qualifications} that temperature just is average kinetic energy, giving the impression that temperature is exact (because average kinetic energy is exact).  The oversimplification is harmless for all practical purposes. However, in our case  the qualifications matter.  In fact, temperature is vague, even when we use a precise number such as  $T=273.15$K. Moreover, it is overdetermined that it is vague. 

According to  kinetic theory of gas,  temperature has a microscopic meaning. For example, the temperature of an ideal gas in equilibrium is proportional to its average (translational) kinetic energy. In symbols:

\begin{equation}\label{temp}
	\bar{E} = \frac{3}{2} k_B T_k 
\end{equation}
where $\bar{E}$ represents the average kinetic energy of the gas molecules, $k_B$ is the Boltzmann constant, and $T_k$ represents the thermodynamic temperature of the gas. Assuming that the collection of gas molecules is an exact notion, and that each molecule has an exact value of (translational) kinetic energy, then the average kinetic energy of the gas is an exact quantity, which equals the sum of  kinetic energies divided by the number of molecules. The constant $\frac{3}{2}$ is obviously exact. If $k_B$ has an exact value, then $T_k$ also has an exact value, for the ideal gas at equilibrium. In this case, for certain ideal gasses in equilibrium, they will have the exact temperature $T=273.15$K. 

However, vagueness enters from at least two sources: (1) the Boltzmann constant and (2) thermal  equilibrium.  The upshot is that having temperature $T=273.15$K is vague and admits borderline cases: for some  gasses in the world, it is not determinate whether they are in the macroscopic state of having temperature $T=273.15$K.

 \begin{figure}
\centerline{\includegraphics[scale=0.35]{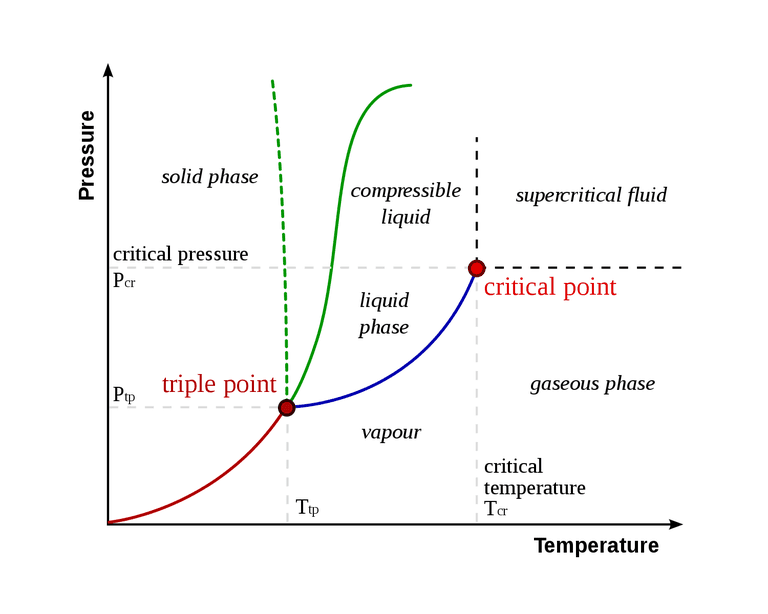}}
\caption{A phase diagram of the triple point of water. \y{Picture by Matthieumarechal, CC BY-SA 3.0 <http://creativecommons.org/licenses/by-sa/3.0/>, via Wikimedia Commons.}}
\end{figure}

First,  the Boltzmann constant, $k_B$, does not seem to have an exact value known to nature (unless we commit to untraceable arbitrariness to be explained in \S3.3). $k_B$ is a physical constant different from those that occur in the dynamical laws, such as the gravitational constant $G$ in $F=G m_1 m_2 / r^2$. Unlike $G$ in the law of gravitation, $k_B$ is a \emph{scaling constant}, playing the role of bridging the microscopic scales of molecules and the macroscopic scales of gas in a box. In this sense, $k_B$ is like the Avogadro number $N_A$. Just as there is no sharp boundary between the microscopic and the macroscopic, there is no sharp boundary between  different values of any macroscopic variable. Historically, $k_B$ is a measured quantity with respect to the triple point of water, a particular state of water where the solid, liquid, and vapor phases of water can coexist in a stable equilibrium (see Figure 3). The triple point also serves as a reference point for $T=273.15$K. But the picture is highly idealized and assumes a sharp transition that would naturally stand for \emph{the triple point}.  In fact, for any body of water in the real world, there is no single \emph{point} that is aptly named the ``triple point.'' At best, there is a quick but  smooth transition that only becomes a sudden jump in the infinite limit (e.g. as the number of particles goes to infinity), which does not obtain in the real world.
 \x{Given the smoothness of the transition, there is no sharp boundary to draw between  states that count as $T=273.15$K and  states that do not,  between states that count as $T=273.15$K and  states that count as borderline $T=273.15$K, and so on.}  And the same issue likely carries over to any phase transition or critical point we encounter. That is the first source of vagueness.\footnote{One could of course stipulate an exact value for  $k_B$. This is actually done recently, at the 26th meeting of the General Conference on Weights and Measures, to define $k_B$ with an exact value instead of referring to it as a measured quantity. See www.bipm.org/en/CGPM/db/26/1. We should understand the redefinition as a practical instruction for conventionally setting international standards for  measurements and calculations. But as for the constant $k_B$ known to Nature, if it has an exact value, it would contain untraceable arbitrariness. See \S3.3. } 

Second, the notion of thermal equilibrium is not exact. To apply Equation (\ref{temp})  and calculate the temperature of some  real gas in front of us, we need to  adjust the equation and account for any differences between  the ideal gas law and the real gas. Suppose that can be done without introducing any additional vagueness. To apply the corrected equation, it still needs to be the case that the gas is in thermal equilibrium. %Assuming (for the sake of argument) all constants in the equation have exact values, and assuming any gas has an exact average kinetic energy, then any gas that is in thermal equilibrium has an exact temperature. It seems to imply that the macrostate described by temperature $T=273.15$K would have an exact boundary of cases including all and only those with certain exact average kinetic energy $\bar{E}_T$: no real gas in the world would be its borderline cases. 

However,  some gasses are borderline cases of being in thermal equilibrium.  For any gas in a box, thermal equilibrium is the ``most likely'' state. It is a state that (roughly) requires that the positions of the gas molecules to be evenly distributed in the box and their velocities conform to a particular Gaussian distribution (the Maxwell-Boltzmann distribution). However, the uniform distribution in positions and the Gaussian distribution in velocities  obtain  in the infinite limit (as the number of gas molecules goes to infinity) and almost never in the real world (when the gas only has a finite number of molecules).  For example, a gas of 100 billion molecules almost never has exactly 50 billion on the left half of the box and 50 billion on the right half, just as an unbiased coin  flipped 100 billion times almost never produces exactly 50 billion heads and exactly 50 billion tails (it only approaches 50/50 as the number of flips tends to infinity). So, to avoid making equilibrium an ``unlikely state,''  we accept the modification that a gas is in equilibrium if its gas molecules are \emph{more or less} uniformly distributed in positions and \emph{more or less} of the Gaussian distribution in velocities. Hence, a gas of 100 billion molecules can be in equilibrium when exactly 50 billion is in the left half of the box and exactly 50 billion is in the right half of the box; but it can stay in equilibrium if there is one more molecule on the left and one fewer on the right; and it can still stay in equilibrium if there are two more on the left and two fewer on the right; and so on.\footnote{To fully describe thermal equilibrium, we need to coarse-grain \emph{more finely} into smaller cells than just two halves, and we need to consider momentum degrees of freedom. But the point made above easily generalizes.} But when does the gas stop being in equilibrium and start being in non-equilibrium? What is the exact meaning of ``more or less'' in the modified definition? We can use the strategy in \S2.1 and run a \emph{sorites argument} here similar to the one on ``bald.''  Hence, there are real gasses in the world that have the required average kinetic energy $\bar{E}_T$  but nonetheless are borderline cases of thermal equilibrium. Thus, such gasses are  borderline cases of having temperature $T=273.15$K.\footnote{A moment's reflection suggests that invoking ``local equilibrium'' does not help; it faces the same problem of vagueness.}  \x{Moreover, there are borderline borderline cases. Suppose a gas with 50 billion - 20 million molecules on the left and 50 billion + 20 million on the right is a borderline case of thermal equilibrium. When do borderline cases begin and when do they end? If we move molecules from the right to the left one by one, eventually we reach determinate thermal equilibrium. However, it is implausible that there is a sharp transition from borderline thermal equilibrium to determinate thermal equilibrium. Hence, there are borderline borderline  cases.  The argument iterates, yielding borderline borderline borderline cases and so on. That is higher-order vagueness.}

%For a  gas that is not in thermal equilibrium, we may divide the gas into smaller parts where each part is a 1x1x1-cubic-inch cell that may instantiate ``local equilibrium.'' Even though the entire box of gas is not in equilibrium, perhaps the gas in each cell is in equilibrium so that we can apply some version of Equation (\ref{temp}) to calculate the local temperature in each cell. But \emph{being in local equilibrium}  is still vague, for the same reasons we discussed before. What if we divide the box into microscopically fine-grained cells such that each box only contains at most one molecule? Would that give us an exact notion of local equilibrium? No, it will not work. Thermal equilibrium is essentially a macroscopic notion. No single gas molecule can be in thermal equilibrium, because the notion applies only when we have a collection of large number of molecules. If the cell contains a large number of molecules, we can apply the notion of equilibrium but that will have vagueness. If the cell contains only one molecule, the notion of equilibrium is not applicable. Similarly, temperature is defined macroscopically, for a large collection of molecules. We can stipulate some microscopic definition of temperature but it is unclear how it connects with the macroscopic notion we use in scientific practice. 

Since neither the Boltzmann constant nor thermal equilibrium is exact, it is overdetermined that the macrostate of  $T=273.15$K  is vague. The same  goes for any other particular temperature, and any particular level of entropy, pressure, and so on.   We have similar reasons to think that almost every other macroscopic variable, as used in thermodynamics and scientific practice, is vague. Hence, we have good reasons to think that WPH is vague: there are some worlds whose initial conditions  are borderline cases  of  macrostate $M_0$.

 \begin{figure}
\centerline{\includegraphics[scale=0.24]{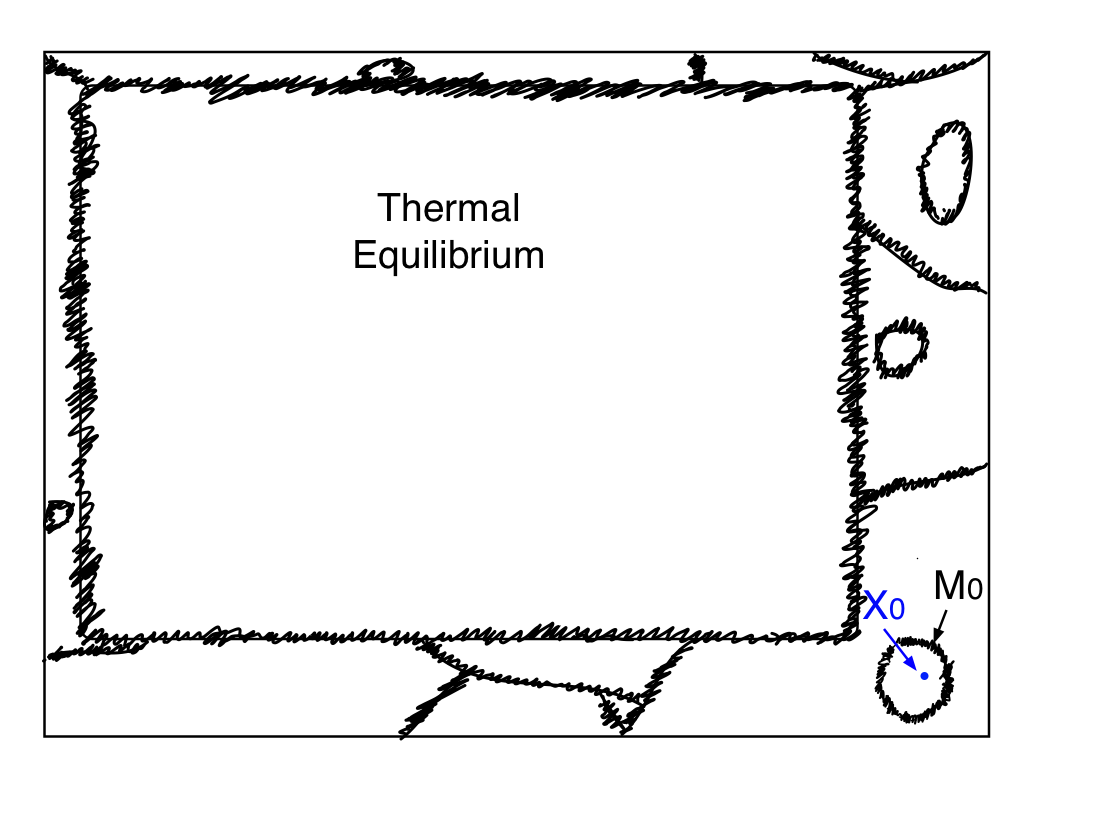}}
\caption{A diagram of  phase space where macrostates have fuzzy boundaries. The macrostate $M_0$ represents the initial low-entropy condition described by WPH. $X_0$ is the actual initial microstate. The picture is not drawn to scale. }
\end{figure}

There is a more systematic way to think about the vagueness of the thermodynamic macrostates in general and the vagueness of $M_0$ in the WPH. In the Boltzmannian account of  classical statistical mechanics, macrostates and microstates can be understood as certain structures on  phase space (Figure 4). 
\begin{itemize}
	\item Phase space: in classical mechanics, phase space is a 6N-dimensional space that encodes all the microscopic possibilities of the system. 
	\item Microstate: a point in phase space, which is a maximally specific description of a system. In classical mechanics, the microstate specifies the positions and the momenta of all particles.  
	\item Macrostate: a region in phase space in which the points inside are macroscopically similar, which is a less detailed and more coarse-grained description of a system. The largest macrostate is thermal equilibrium.
	\item Fuzziness: the partition of phase space into macrostates is not exact; the macrostates have fuzzy boundaries. Their boundaries  become exact only given some choices of the ``C-parameters'', including the size of  cells for coarse-graining and the correspondence between distribution functions and macroscopic variables. 
	\item Entropy:  $S(x) = k_B \text{log} |M(x)|,$ where $| \cdot |$ denotes the standard volume measure in phase space. Because of Fuzziness, in general, the (Boltzmann) entropy of a system is not exact. 
\end{itemize} 
We can translate WPH into the language of phase space: at one temporal boundary of space-time, the microstate of the universe $X_0$ lies inside  a particular macrostate $M_0$ that has low volume in phase space. This shows that the derived properties invoked by WPH can be re-expressed in terms of the fundamental properties (volume of microstates in some region of phase space, which can be further expressed as Lebesgue measure of sets of ordered 6N-tuples of positions and momenta of particles). Unlike properties such as tigerhood, and more like  acceleration,  thermodynamic properties such as temperature and entropy are  more natural and more easily re-expressed in terms of fundamental properties. Hence, the re-expressed version of WPH can satisfy even the strict criterion that  only allows fundamental properties to appear in fundamental laws.

\x{Fuzziness is crucial for understanding the  vagueness and higher-order vagueness of  macrostates.   Without specifying the exact values (or exact ranges of values) of the C-parameters, the macrostates  have fuzzy boundaries: some microstates are borderline cases for certain macrostates, some are borderline borderline cases, and so on. The fuzzy boundary of $M_0$ illustrates  the existence of borderline microstates and higher-order vagueness. } There will be a precise identification of macrostates with sets of microstates only when we exactly specify the C-parameters (or their ranges). In other words, there is a precise partition of microstates on phase space into regions that are macroscopically similar (macrostates) only when we make some arbitrary choices about what the C-parameters are. In such situations, the WPH macrostate $M_0$ would correspond to an exact set $\Gamma_0$ on phase space, and the initial microstate has to be contained in $\Gamma_0$.

However, proponents of the WPH do not  specify a precise set. A precise set $\Gamma_0$ would require more precision than is given in statistical mechanics---it requires the specific values of the coarse-grained cells and the specific correspondence with distribution functions. (In the standard quantum case discussed in Appendix, it also requires the precise cut-off threshold for when a superposition   belongs to a macrostate.)  The precise values of the C-parameters could be added to the theory to make WPH into a precise statement (which we call the \textit{Strong Past Hypothesis} in \S3.3). But they are nowhere to be found in the proposal, and rightly so.\footnote{For example, see descriptions of SWPH and WPH in \cite{goldstein2001boltzmann}, \cite{albert2000time}, and \cite{carroll2010eternity}. }

Some choices of the C-parameters are clearly unacceptable. If the coarse-graining cells are too large, they cannot reflect the  variations in the values of macroscopic variables; if the coarse-graining cells are too small, they may not contain enough gas molecules to be statistically significant. Hence, they have to be macroscopically small but microscopically large (\cite{albert2000time} p.44(fn.5) and \cite{goldstein2019gibbs}).  However, if we were to make the parameters (or the ranges of parameters) more and more precise, beyond a certain point, any extra precision in the choice would seem completely arbitrary. They correspond to how large the cells are and which function is the correct one when defining the relation between temperature and sets of microstates. That does not seem to correspond to any objective facts in the world. (How large is large enough and how small is small enough?)  In this respect, the arbitrariness in precise C-parameters is quite unlike that in the fundamental dynamical constants. (In \S3.3, we discuss their differences in terms of a theoretical virtue called `traceability.') Moreover, not only do we lack precise parameters, we also lack a precise set of permissible parameters (hence no exact ranges of values for the C-parameters). There shouldn't be sharp boundaries anywhere. \x{Suppose size $m$ is borderline large enough and size $n$ is determinately large enough. Small changes from $m$ will eventually get us to $n$, but it is implausible that there is a sharp transition from borderline large enough to determinately large enough. Similarly, there shouldn't be a sharp transition between borderline large enough to borderline borderline large enough, and so on.  That is higher-order vagueness.}

Because of higher-order vagueness, \x{we need to take standard mathematical representation of WPH with a grain of salt}. The macroscopic variables---adjustable parameters in WPH---need to be coarse-grained enough to respect the vagueness. For example, we may \textit{represent} the temperature of $M_0$ as $10^{32}$ degrees Kelvin. But temperature does not have the exactness of real numbers. A more careful way to represent the vague temperature should be ``$10^{32}$-ish degrees Kelvin,'' where the ``-ish'' qualifier signifies that temperature is vague and the number $10^{32}$ is only an imperfect mathematical representation.\footnote{Thanks to Alan H\'ajek for discussions here.} Its  exactness is artificial. Hence, WPH should be characterized as a macrostate $M_0$ specified by  $S_0$-ish entropy, $T_0$-ish temperature, and so on. 

The vagueness here is appropriate, since  macroscopic variables only make sense when there are enough degrees of freedom (such as a large number of particles). In practice, however, such vagueness rarely matters:  there will be enough margins  such that to explain the thermodynamic phenomena, which are themselves vague, we do not need the extra exactness. The vagueness disappears \emph{for all practical purposes}. Nevertheless,  WPH is a genuine case of   \x{fundamental} nomic vagueness and it is a possibility to take seriously.

\subsection{Untraceable Arbitrariness of the Strong Past Hypothesis}

%We suggested that the vagueness of the Past Hypothesis should not be considered as epistemic, for otherwise it would impose an objectionable kind of arbitrariness to the world akin to the arbitrariness of a privileged coordinate system. However, we note that it is different from the arbitrariness of having exact values for the natural constants. We make this point more precise in this section. 

For the sake of completeness, I consider the possibility of an exact version of PH. Suppose there is an exact law known to nature and the vagueness of WPH is only epistemic: there is, in fact, a precise set $\Gamma_0$ with exact boundaries on phase space that stands in for the initial macrostate. %Regardless of one's view about nomic vagueness and one's favorite theory of lawhood, it may be helpful to consider the pros and cons of such a proposal as a scientific postulate independent of one's metaphysical inclinations. 
I formulate it as follows: 
\begin{description}
	\item[Strong Past Hypothesis (SPH)] At one temporal boundary of space-time, the microstate of the universe is in $\Gamma_0$, where $\Gamma_0$ corresponds to a particular admissible precisification of $M_0$. 
\end{description}
Unlike WPH, SPH is exact. As such, it is mathematically expressible. However, as I explain below, SPH violates a plausible feature that every other fundamental law and dynamical constant satisfies: SPH is ``untraceable.'' The exact boundary of $\Gamma_0$ does not ``leave a trace'' in typical worlds compatible with it. Hence, SPH  is arbitrary in a way that other exact fundamental postulates in physics are not. Moreover, it widens the gap between the ontic and the nomic. Other things being equal,  that seems to make it less appealing among  proposals for understanding PH as a fundamental law. 

On the epistemic interpretation of vagueness, there is in fact an  exact number of hairs, $n$, that turns someone from being bald to being non-bald. But the number $n$ is not known to us. In fact, it cannot be known to us in any way. Similarly, there are in fact  exact boundaries of the macrostate $M_0$, represented by the set of microstates $\Gamma_0$ on phase space. The exact set can be picked out only by the unhelpful description ``the set that is invoked by the SPH.'' Which set it is is unknown and likely unknowable by empirical investigations (as I explain below). However, many things that are true of nature may be unknown or unknowable to us, as a consequence of certain physical laws. There are examples of in-principle limitations of knowledge in well-defined physical theories such as Bohmian mechanics and GRW collapse theories  (\cite{cowan2016epistemology}). Moreover, we may not know the exact values of the fundamental constants and the fundamental dynamical laws, if not forever then at least for a long time. Hence, knowledge and knowability about the precise boundaries of $\Gamma_0$ is not the issue, for they may also arise for other fundamental laws and dynamical constants that we think are fine. Neither is the problem that the postulate of a precise set $\Gamma_0$ would be a brute fact that is not explained further. Every other fundamental law or dynamical constant is supposed to be also brute and not explained further (in the scientific sense and not in the metaphysical sense of explanation). 

%The problematic nature of the exactness of SPH is closer to that of a privileged coordinate system of Euclidean space $\mathbb{E}^3$. It would be extremely surprising if   there were in fact a privileged coordinate system of $\mathbb{E}^3$ known to nature, not because it would be unknowable  but because the choice would seem completely arbitrary. We usually choose some coordinate system for convenience. The choice does not seem to correspond to any worldly facts.  Hence, we have strong reasons to reject a ``Strong Euclidean Hypothesis'' that postulates a privileged coordinate system of space. Similarly, considering SPH as a fundamental postulate of physics, we would have  reasons to be suspicious. When we choose any particular  C-parameters, we do it for convenience or in accordance with some arbitrary convention. It is hard to believe any such choice would be preferred by the nomological structure of the world. 

What sets the arbitrariness of SPH apart from that of the dynamical constants and other fundamental laws is its \emph{untraceability}. The exact boundaries  of  SPH are typically untraceable.  There are infinitely many ways to  change the boundaries of $\Gamma_0$ that do not lead to any  differences for most worlds SPH deems possible. Hence, $\Gamma_0$ does not leave a trace in most worlds compatible with it. %That would be a striking feature, since no other fundamental laws or dynamical constants are similarly untraceable. 

Let me make this notion of traceability more precise. It is plausible that the objective features of the world are  reflected in the changes in the properties of particles, the field configurations, the mass densities, the space-time geometry, and so on. Such changes do not have to be measured or measurable by human beings. But  for the familiar fundamental laws and their dynamical constants, typical changes in their exact values will be ``felt'' by the matter distributions (or some other part of the fundamental ontology excluding the fundamental laws) in the nomologically possible worlds.  That is, there  are some worldly features in the fundamental  material ontology that are sensitive to typical changes in the ``nomology.'' For example, any changes in the gravitational constant $G$ will be felt by the massive objects and will change (however slightly or significantly) the motion of planets around stars, the formation of galaxies, and the distribution of fundamental matter. On a closer scale, it affects how exactly my vase shatters when it hits the ground. In other words, there should be \emph{some traces in the material ontology} of the world. If the value of $G$ had been different, the material ontology would have been different.  We can capture this idea modally as  changes in the nomological status (from possible to impossible) or the objective (conditional) probability (e.g. from 0.8 to 0.3 given prior histories) of the world. We formulate the following condition on traceability: 

\begin{description}
	\item[Traceability-at-a-World] A certain adjustable parameter $O$ in the physical law $L$ is \emph{traceable at world $w$} if any change in $O$ (while holding other parameters fixed) will result in some change in the nomological status of $w$ with respect to $L$, i.e. from possible to impossible or from likely to unlikely (or some other change in the probabilistic measures). 
	\end{description}
  We are treating ``adjustable parameter'' in a loose sense.	For example, in the case of Newtonian theory with gravitation $F_G=ma \land F_G =G m_1 m_2/r^2$, we can adjust it in the following (independent) ways:
\begin{itemize}
	\item Change the constant  G= 6.67430 to G'=6.68 (in the appropriate unit);
	\item Change division by $r^2$ to division by $r^{2.001}$;
	\item Change the multiplication by  $m_1$ to multiplication by $m_1^{1.00001}$.
\end{itemize}
All such changes are \emph{traceable at typical worlds} that satisfy Newton's law of motion and law of universal gravitation. For a typical Newtonian world whose microscopic history $h$ is a solution to the Newtonian laws, $h$ will not be possible given any of those changes. In other words, it will change a typical history $h$ from nomologically possible to nomologically impossible with respect to Newtonian theory of gravitation. Here, we are interested in traceability at most worlds that are allowed by  $L$. This is because there may be cases where for ``accidental'' reasons two different values of $O$ may produce the same world $w$ in exact microscopic details; a change of the value of $O$ does not change $w$ from possible to impossible or change its probability. Such cases would be atypical. The relevant property is this:
\begin{description}
	\item[Traceability] A certain adjustable parameter $O$ in the physical law $L$ is \emph{traceable} if  	$O$ is traceable at most worlds allowed by $L$.\footnote{Here ``most'' is with respect to some natural measure in the state space such as the Lebesgue measure in phase space or normalized surface area measure on the unit sphere in Hilbert space. The threshold for most  is vague. As such, there may be borderline cases of traceable parameters. This is to be expected, as traceability is supposed to be a theoretical virtue; like other theoretical virtues, it can be vague. But the examples we encounter here are  clear-cut. }
	\end{description}
	% In fact, many of the familiar cases satisfy a stronger criterion: the adjustable parameters in the physical theory are traceable at \emph{typical} worlds. 
	 If some degree of freedom $O$ is  traceable at most worlds, then at most worlds (typically)  the value of $O$ can be determined  to arbitrary microscopic precision. Then, normally, the more information we know about the actual world the more precise we can determine the value of $O$. However, what matters is not our epistemic access. For typical worlds compatible with $L$, if it is deterministic, most  worlds will only admit one value of $O$. That is the case for the gravitational constant $G$.
	 
Similarly, the laws and dynamical constants of Maxwellian electrodynamics are traceable; those of Bohmian mechanics are traceable; those of Everettian quantum theory are traceable; those of special and general relativity are traceable. In those theories, there is a tight connection between the nomic and the ontic. Typically the precision of the laws leaves traces in the material ontology. 
%\footnote{Some toy physical theories may violate traceability but the violations usually involve highly atypical changes in some parameters, such as the changes from using real-valued quantities to irrational-valued quantities. We set them aside for the purpose of this paper.} 

A stochastic theory such as GRW presents an interesting wrinkle. The GRW theory postulates two fundamental constants: the collapse rate $\lambda$ and the collapse width $\sigma$. Consider just the collapse rate $\lambda$ that describes the probability of collapse (per-particle-per-unit-time). Since it is a probabilistic theory, the same history of quantum states can be compatible with distinct values of $\lambda$. What $\lambda$ does is to provide a probability measure (together with a slightly-modified Born-rule probability measure) that tells us which collapse histories are typical (or very probable) and which are not. However, each  micro-history receives zero measure. It is the macro-history (considered as a set of micro-histories that are macroscopically similar) that can receive positive  probabilities. Hence, in a stochastic theory, we should understand the appropriate change of nomological status as changes in the probabilistic measure of the macro-history that the micro-history realizes. 

 \begin{figure}
\centerline{\includegraphics[scale=0.24]{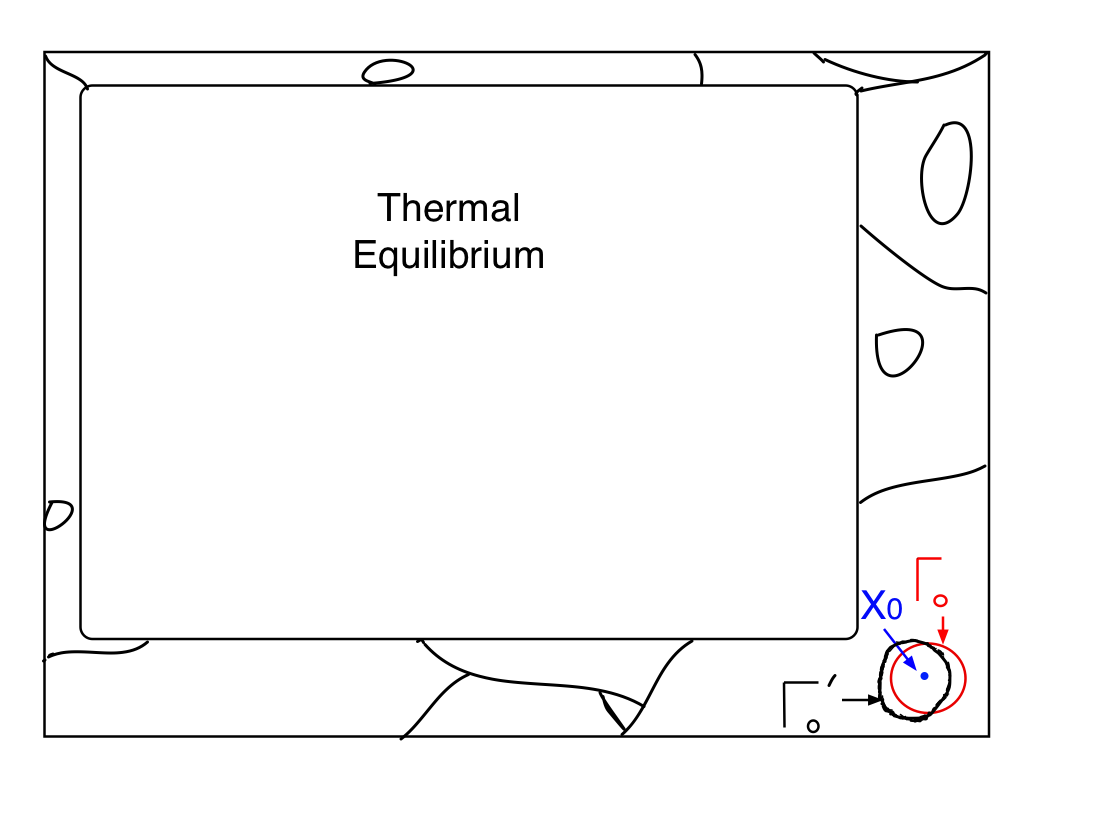}}
\caption{ A diagram of phase space where macrostates have exact boundaries. $\Gamma_0$ and $\Gamma_0'$ are two admissible precisifications of $M_0$. The actual initial microstate $X_0$ lies inside both. }
\end{figure}

Although familiar laws in physics are traceable, SPH is not. To see this, consider $\Gamma_0$ and another set $\Gamma_0'$ that has slightly different boundaries (see Figure 5). Suppose both are admissible precisifications of $M_0$ and both include the actual initial microstate $X_0$ as a member. Then the world starting in microstate $X_0$ is compatible with SPH and another law SPH' that slightly alters the boundaries of $\Gamma_0$. Moreover, this is the case for typical worlds compatible with  $\Gamma_0$: at most worlds compatible with $\Gamma_0$, slightly altering the boundaries of $\Gamma_0$ will not make a difference to the nomological status of most worlds. (At some atypical worlds very close to the boundaries of $\Gamma_0$, altering the boundaries will take them from being possible to being impossible or vice versa.) For most worlds inside $\Gamma_0$, there will be infinitely many changes to the boundaries of $\Gamma_0$ that do not affect the probabilities of those worlds.

Hence, SPH is not traceable.\footnote{Because of higher-order vagueness, the same will be true for a disjunctive version of SPH that says that the initial microstate belongs to  a determinate set of precisifications, such as: $X_0$ is in $ \Gamma_0$ or $\Gamma_0'$ or $ \Gamma_0''$ or $ \Gamma_0'''$. } And there lies the key difference between SPH and other fundamental laws and constants.  The former is arbitrary in a way the latter are not: SPH has untraceable arbitrariness. For traceable laws and constants such as $G$, their  values may be arbitrarily precise; their values are not explained further. However, they still respect a close connection between the nomic and the ontic: their exact values are reflected in the material ontology. That is not the case of SPH; the exact boundaries of $\Gamma_0$ outrun the ontic; the exact choice of $\Gamma_0$ is not reflected in the material ontology. \emph{Other things being equal, we should minimize the gap between the ontic and the nomic.} (To emphasize: this is different from the gap between  the nomic and what's epistemically accessible, for plenty of facts about the material ontology may forever lie beyond our epistemic ken.)

It is implausible that we can appeal to super-empirical virtues to pin down $\Gamma_0$. Take for example the theoretical virtue of simplicity. It is unlikely that there will be a simplest precisification of $M_0$, just as it is unlikely there is a simplest choice of the coarse-graining size (or other C-parameters).\footnote{\cite{penrose1979singularities}'s  Weyl curvature hypothesis (WCP) is a simple and exact version of the low-entropy initial condition in classical relativity. But by itself it seems insufficient to explain the records asymmetry \citep{rovelli2020memory}. Moreover, the quantum generalization of WCP \citep{ashtekar2016initial} is vague.} Furthermore, those theoretical virtues are themselves vague. In cases where Nature is kind to us, there may be a choice that is by far the simplest (or best balances various virtues) that their vagueness makes no difference. However, although we may have \emph{faith} in Nature's kindness, we have no \emph{reason} to think that SPH is such a case. 

Endorsing SPH leads to endorsing some untraceable arbitrariness in Nature.
%\footnote{Traceability is different from  Lewis's thesis of Humean supervenience (HS): no differences in the nomology without there being some differences in the mosaic.   For example, they do not imply each other. For one direction, we note that  it is logically possible that Humean supervenience is true and the best summary of the mosaic fails traceability. For the other direction, we note that it is logically possible that there are non-Humean GRW-type laws---GRW1 and GRW2---that are stochastic and have different collapse rates $\lambda_1$ and $\lambda_2$. Since they are non-Humean, they can obtain with the same mosaic.  Yet the collapse rate  can still be  traceable: different choices of  the collapse rate can assign different probability measures over typical worlds (their macro-histories). }  
Although it is not impossible Nature acts in this strange way, if every other fundamental postulate and dynamical constant in physics seems to respect traceability, conservativeness suggests that we  try to keep it if we can. We should respect the tight connection between the nomic and the ontic by not letting in untreaceable arbitrariness. 

Is WPH also  untraceably arbitrary? No. Let me  explain in what sense  WPH is traceable. WPH (see Figure 4) does not delineate the nomological possibilities as a set of microstates with sharp boundaries. Instead, it selects a fuzzy region---$M_0$.\footnote{Here it may be useful to follow \cite{sainsbury1990concepts}'s suggestion: we should not be speaking of ``boundaries'' at all, for there is only one kind of boundary. Hence, a fuzzy region is perhaps a better mental imagery.}  WPH is  stated in the vague macro language: the initial macrostate has $S_0$-ish entropy, $T_0$-ish temperature, and so on. The vagueness of the ``-ish'' tolerates small differences in values: ``$20$-ish'' and ``$20.0001$-ish'' are \textit{equivalent representations} of the same macrostate. Thus, slight variations of WPH will not produce  different vague laws.  Adjusting the entropy value from $S_0$ to $S_0+\epsilon$ will select the same fuzzy region, as long as $\epsilon$ is small enough. Moreover, increasing  entropy by  too much will produce  observable, macroscopic differences in most worlds compatible with WPH (e.g. from low entropy to high entropy). Of course, what counts as small enough and what counts as too much will be vague.  Now,  even though not all traceable differences are observable, observable differences are traceable. Hence, unlike the exact SPH,  the vague WPH is not untraceably arbitrary.

Interestingly, although traceability  seems like a novel theoretical virtue, it provides another explanation for people's  negative attitudes towards the quantum measurement axioms.  Many philosophers of physics are unfriendly towards  fundamental yet vague quantum measurement axioms, but some of them can accept a fundamental WPH. Both are vague. What can be a principled reason that distinguishes the two cases? There are in fact two reasons. First, there is the ideal of unification mentioned in \S2.5.  The second reason, which has  so far been under-appreciated, is that the exact alternative of WPH is untraceable, while  the exact alternative of the measurement axioms is in fact \emph{traceable}; different cut-offs in the law will typically lead to differences in the fundamental material ontology. A particular simple alternative to the vague measurement axiom is provided by the GRW theory. Exact values of the GRW adjustable parameters (collapse width, collapse rate, and probability of the collapse center)  leave traces in the material world.  

It is difficult to formulate precise theory-choice principles, but here is a proposal: \textit{other things being equal, if we can avoid  \x{fundamental} nomic vagueness without introducing untraceable arbitrariness, we should prefer an exact alternative; but if we can do it only by introducing untraceable arbitrariness, then a fundamental yet vague law is perfectly acceptable and should be preferred to the exact alternative.} Given the proviso  ``other things being equal,''  the principle should be applied very carefully. We should treat  \x{fundamental} nomic vagueness as a defeasible indicator that can, in some cases, reveal the deeper defect in the theory. For example, the vagueness of textbook quantum mechanics indicates that the theory lacks  a unified explanation, where the precise alternative GRW  not only ensures exactness but also provides ``unified dynamics for microscopic and macroscopic systems'' \citep{ghirardi1986unified}. However, since the indicator is \textit{defeasible},  some theories  can contain  \x{fundamental} nomic vagueness and yet be perfectly fine as  candidate fundamental theories, such as the case of the classical statistical mechanics with WPH (the Mentaculus). In that case,  \x{fundamental} nomic exactness is not worth the price, for it would make the theory untraceably arbitrary;  \x{fundamental} nomic vagueness is not a symptom of some deeper disunity in the Mentaculus, as the theory aspires to be a unified explanation of the fundamental and the non-fundamental sciences.

%explains our different attitude towards the quantum measurement axioms (\S2.5).  Many philosophers of physics are unfriendly towards  fundamental yet vague quantum measurement axioms, but some of them will be much happier with a fundamental WPH. Both are vague. What can be a principled reason that distinguishes the two cases? In the case of WPH, its exact alternative (SPH) with precise boundaries is untraceable. In the case of the  quantum measurement axioms, their exact alternative is in fact \emph{traceable}: different cut-offs in the law will typically lead to differences in the fundamental material ontology. All else being equal, if we can avoid nomic vagueness without committing untraceable arbitrariness, we should prefer an exact alternative. But if we can do it only if we commit untraceable arbitrariness, then a fundamental yet vague law is perfectly acceptable.  This provides another principled reason to prefer Bohm's theory and GRW theory over textbook quantum mechanics, even if one does not subscribe to the unity of nature ideal. 

I have not provided a decisive argument for the theory-choice principle. For people desiring to avoid  \x{fundamental} nomic vagueness at all costs, they can reject the principle and choose SPH over WPH as the fundamental boundary-condition law.  
%Of course, conservativeness in this regard should be balanced with conservativeness in another dimension: we might hope that all fundamental laws are exact, and SPH would eliminate nomic vagueness. 
Therefore, a more neutral way to summarize the findings so far is that we face a dilemma: either accept vague fundamental laws such as SWPH and WPH, or violate traceability by adopting SPH. There is no free lunch in Nature; either way we have to pay. 

\section{\x{Fundamental} Nomic Exactness without Untraceable Arbitrariness}

Is the dilemma an essential feature in any theory that includes a fundamental law of a low-entropy boundary-condition? In this section, I consider under what conditions the dilemma may be dissolved. It turns out that the dilemma is essential if our world is classical but dissolvable if our world is quantum. Here I focus on the conceptual elements and leave the technical details to the Appendix. 

The dilemma arises for PH because of its role in the fundamental physical theory.  In the Mentaculus, the micro-dynamical laws (such as $F=ma$) suffice to explain the entire history of every microstate, and PH is brought into the theory to explain why most nomologically possible microstates should display the macroscopic time-asymmetric regularities, such as those summarized in the Second Law.  It does so by ``throwing out'' most microstates that do not display the regularities. However, the explanandum---the time-asymmetric regularities---are vague, macroscopic phenomena about temperature, entropy and the like. Hence, a vague law such as WPH would suffice as the explanans and an exact law such as SPH would be overly precise. In contrast, micro-dynamical laws are traceable; their explanandum are exact, microscopic phenomena (e.g.  the relative distances of point particles). This observations leads to an interesting possibility: if the low-entropy boundary condition somehow appears in the micro-dynamical equations, perhaps an exact version of PH will no longer be untraceably arbitrary.  

In \S3.3, SPH is formulated as a constraint on $X_0$, the initial classical state of the universe, which represents the positions and momenta of all the particles. The $\Gamma_0$ stipulated by SPH does not appear in the traceable dynamical equation  $F=ma$.  In realist theories of quantum mechanics, SPH can be formulated as a constraint on the initial quantum state of the universe, represented by a wave function, $\Psi_0$. Still, the exact boundary stipulated by SPH does not appear in the traceable dynamical equation (the Sch\"odinger equation). As such, the quantum version of SPH contains the same vice as the classical one---untraceable arbitrariness. Thus, we can similarly argue that we should prefer the quantum WPH over the quantum SPH. 
 
However, quantum theory contains a more general kind of quantum states, represented by density matrices. It has been recognized for many years that the fundamental quantum state of the universe may be in such a state, which we shall call $W$, and it would have the same observable consequences as one represented by $\Psi$.  In parallel with the realist quantum theories with a fundamental $\Psi$, there have been developments of $W$-versions of Bohmian mechanics where $W$ directly guides Bohmian particles, GRW collapse theories where $W$ undergoes spontaneous collapses, and Everettian theories where $W$ strictly obeys the von Neumann equation and realizes an emergent multiverse.  The viability of the $W$-theories as an alternative to $\Psi$-theories  transforms the situation about  \x{fundamental} nomic vagueness and untraceable arbitrariness in a quantum world. It is good news for the dissolving the dilemma, but this strategy has so far been under-appreciated (likely because   \x{fundamental} nomic vagueness is under-appreciated).   

Here is the innovation allowed by $W$-theories:  the fundamental quantum state of the universe is initially $W_0$ (a density matrix), and we further stipulate that $W_0$ is \textit{the} simplest quantum state (the normalized projection) corresponding to a particular precisification of a low-entropy macrostate (represented by $\mathscr{H}_{PH}$). In \cite{chen2018IPH, chen2018HU}, \y{I call this} the \textit{Initial Projection Hypothesis} (IPH). IPH is a new proposal of the fundamental law of a low-entropy boundary condition that replaces (quantum) SPH. \y{I call the package of IPH and fundamental dynamical laws \textit{the Wentaculus}.} Unlike SPH and WPH in standard $\Psi$-quantum theories, IPH selects a unique quantum state---it is stipulated to be both the actual one and the only nomologically possible one. (Yet, IPH remains simple insofar as $\mathscr{H}_{PH}$ is simple.)  IPH is exact but also traceable. It is exact because, unlike WPH, IPH admits no borderline worlds; slight changes of $W_0$ will produce different laws. It is traceable because, unlike SPH,  slight differences in $W_0$ will leave traces in the material ontology. The dilemma is dissolved here because $W_0$  plays two roles at $t_0$: it corresponds to the low-entropy macrostate and also appears in the microscopic equations. It becomes traceable because of its second role, while standard exact versions of PH do not play the second role and  are untraceable.  Moreover, for each realist theory, its $W$-version is no less simple or theoretically virtuous. In fact, one could argue that with IPH, there is a greater degree of unification of  quantum theory with temporal asymmetry. 

The strategy does not work  in a classical world. One could try to replace SPH+SP with a probability distribution $\rho$ (which is analogous to $W$) and somehow use it to guide the motion of classical particles. But it would be awkward: it would either  multiply the ontology to   \textit{particles and the physical counterpart of }$\rho$, or complicates the simple dynamics of $F=ma$ by introducing stochastic jumps. Alternatively, one could completely throw out the classical particles and represent the material ontology with $\rho$. But again, without changing the deterministic dynamics to a stochastic one, one would have to embrace a many-worlds interpretation even in a classical world, because  $\rho$ would deterministically evolve to be supported on different macrostates. All of these changes are artificial. Thus, in a classical world, the  exact alternative to WPH is not worth the price, and a vague WPH is perfectly fine.   The proviso of the theory-choice principle in \S3.3 is not satisfied, as other things are clearly not equal.

\section{Conclusion}

\x{Fundamental nomic vagueness is vagueness in the fundamental laws of nature. On the proposed account, to find out whether a fundamental law is vague, we  check  whether it admits borderline nomologically possible worlds, lacks a well-defined extension, carries sorites-susceptibility, and possesses higher-order vagueness. If the account is intelligible, which I think it is, then fundamental laws of nature is a new place to look  for  vagueness.} The account leaves room for \emph{vague chances}, which is left for future work. \y{I also leave to future work how fundamental nomic vagueness can impact theories of causation, counterfactuals, and scientific explanation.}

\x{On a widespread view where fundamental laws are metaphysically fundamental, fundamental  nomic vagueness is a ``worldly'' kind of vagueness. On such a view, vagueness of fundamental laws implies  vagueness of certain fundamental facts of the world. However, fundamental nomic vagueness differs from standard cases of ontic vagueness, as the latter but not the former  concerns the vague identity, spatio-temporal boundaries, or parts of material objects. }

 It is surprising, whether from  a Humean or a non-Humean perspective, that actual fundamental laws of nature can be vague. We might expect all fundamental laws to be \textit{completely and faithfully expressible} by precise mathematical equations. That expectation will be mistaken if actual fundamental laws include vague ones such as  WPH  \x{and if higher-order vagueness is not completely mathematically expressible}. One can use  SPH to eliminate vagueness by fiat. However, it introduces \emph{untraceable  arbitrariness} that is absent in any other fundamental laws or dynamical constants. 
 
 Interestingly, the dilemma between  \x{fundamental} nomic vagueness and untraceability is dissolved when we directly use the initial macrostate to dictate (or describe) the motion of fundamental material objects, such as when we postulate IPH in density-matrix quantum theories. Hence, another surprising lesson is that, far from making the world vague, the innovations of quantum theory can eliminate  \x{fundamental} nomic vagueness  without introducing  untraceable arbitrariness in Nature. Does it follow we should prefer the density-matrix theories with IPH over the wave-function theories with WPH? That is  an open  question, as we need to carefully consider their strengths and weaknesses,  and whether other things are equal.  \x{Fundamental}  nomic vagueness provides a relevant desideratum for theory-choice. \y{A related question, which I did not fully address in this paper, is the value of preserving complete mathematical expressibility of fundamental physical laws.} 
   
 In current and future physics, there may be other cases of  \x{fundamental} nomic vagueness that cannot be eliminated in a similar manner and cases of  arbitrariness that have a different character. Then one's position on the metaphysics of laws  could make a difference to how one should deal with  \x{fundamental}  nomic vagueness and arbitrariness. As the case study shows, the issue is delicate and should not be settled in advance.  There are many interesting questions  concerning metaphysics, physics, and mathematics that can make a difference to  how much we should tolerate vagueness in the fundamental laws. Attending to those details may also teach us something new about the nature of laws.

\section*{Appendix}

First, I  briefly explain the standard framework of the Boltzmannian account of quantum statistical mechanics (\cite{goldstein2019gibbs}), in parallel to that of the classical statistical mechanics (\S3.2):

 \begin{itemize}
	\item Hilbert space: Hilbert space is a vector space equipped with inner product structure that encodes all the microscopic possibilities (possible worlds) of the system (or the universe as a whole). 
	\item Microstate: a vector in Hilbert space, a maximally specific description of a system.\footnote{It is possible to have additional ontologies such as the Bohmian particles and the GRW mass densities.} 
	\item Macrostate: a subspace in Hilbert space in which the quantum states contained within are macroscopically similar, which is a less detailed and more coarse-grained description of a system. The Hilbert space is orthogonally decomposed into subspaces. 
	\item Fuzziness: the decomposition of Hilbert space into macrostates is not determinate; the macrostates have fuzzy boundaries. Their boundaries become exact only  given some choices of the C-parameters, including the size of  cells for coarse-graining, the correspondence between distribution functions and macroscopic variables, and  the cut-off threshold for macrostate inclusion.
	\item Entropy:  $S(\psi) = k_B \text{dim} \mathscr{H},$ where dim denotes the dimension counting in Hilbert space and $\mathscr{H}$ is the subspace that contains most of $\psi$. Higher-dimensional subspaces tend to have higher entropies. Because of Fuzziness, in general, the (Boltzmann) entropy of a system is not exact. 
\end{itemize}
 We can translate the WPH  in the language of Hilbert space: 
 
 \begin{description}
	\item[Quantum Weak Past Hypothesis (QWPH)] At one temporal boundary of space-time, the wave function of the universe is in a particular macrostate $M_0$, where $M_0$ is the low-entropy macrostate characterized by   the Big Bang cosmology. 
\end{description}
SP would take the following form:
 \begin{description}
	\item[Quantum Statistical Postulate (QSP)] At one temporal boundary of space-time, the probability distribution is the uniform one (with respect to the normalized surface area measure on the unit sphere) over wave functions compatible with QWPH.  
\end{description}
%As in the classical case, both QWPH and QSP are vague laws. Given some choices of C-parameters, the macrostate $M_0$ will correspond to a precise subspace $\mathscr{H}_{PH}$ and the uniform probability distribution will be the surface area measure on the unit sphere in $\mathscr{H}_{PH}$. The actual wave function $\Psi_0$  will be dynamically central, and it basically screens off the dynamical influence of   $\mathscr{H}_{PH}$ on the microscopic histories. So the precision of  $\mathscr{H}_{PH}$ becomes  untraceable at typical worlds (represented by different wave functions).  Hence, a Quantum Strong Past Hypothesis (QSPH) that chooses a particular precise subspace $\mathscr{H}_0$ will be untraceable. Hence, if we stick with these versions of PH, we still have the dilemma between nomic vagueness and untraceable arbitrariness. 

Second, I explain  the new strategy of reconciling \x{fundamental} nomic exactness with traceability. Recent works in the foundations of quantum  mechanics suggest that density matrices can directly represent fundamental quantum states.\footnote{See, for example, \cite{durr2005role, maroney2005density}, \cite{wallace2011logic, wallace2012emergent}, and \cite{chen2018IPH}.} \cite{chen2018IPH} calls the view \emph{density matrix realism}.  On density matrix realism, the fundamental state of the universe is mixed rather than pure and it has to be described by a density matrix rather than a wave function. The density matrix is not merely epistemic.  Since the density matrix is fundamental, the \emph{fundamental micro-dynamics} needs to be revised (e.g.  \`a la \cite{allori2013predictions}) to reflect the change: we  replace the Schr\"odinger equation with the von Neumann equation, the Bohmian guidance equation with another that uses the density matrix as an input, the GRW collapse equations with another that stochastically evolves the density matrix, and various definitions of local beables from the wave function with their density-matrix counterparts. Moreover, the density-matrix versions of Bohm, GRW, and Everett are empirically equivalent to the respective wave-function versions.

I consider a new postulate about the initial quantum state:  
\begin{equation}
	W_0 = \frac{\mathbb{I}_{PH}}{\text{dim} \mathscr{H}_{PH}}
\end{equation}
where $\mathbb{I}_{PH}$ designates the projection operator onto $\mathscr{H}_{PH}$ and dim counts the dimension of that subspace. I use this to formulate a new  low-entropy initial condition for density matrix realism: 
 \begin{description}
	\item[Initial Projection Hypothesis (IPH)] At one temporal boundary of space-time,  the quantum state of the universe is exactly  $W_0$ as described in Equation (2). 
\end{description}
This version of PH is exact. The combination of density matrix realism and IPH does the heavy-lifting. The low-entropy initial condition is completely and unambiguously described in $W_0$. Unlike the SPH or its quantum version QSPH we encountered earlier, $W_0$ enters directly into the fundamental micro-dynamics. Hence, $W_0$ will be traceable and not objectionably arbitrary. To see that , consider  two realist interpretations of the quantum state (\cite{chen2019realism}):
\begin{enumerate}
	\item $W_0$ is ontological: if the initial density matrix represents something in the fundamental material ontology, IPH is obviously traceable. Any changes to the physical values of $W_0$  will leave a trace in every world compatible with IPH. 
	\item $W_0$ is nomological: if the initial density matrix is on a par with the fundamental laws, then $W_0$ plays the same role as the classical Hamiltonian function or fundamental dynamical constant of nature. It is traceable in the Everettian version with a matter-density ontology as the initial matter-density is obtained from $W_0$. It is similarly traceable in the GRW version with a matter-density ontology. For the GRW version with a flash ontology, different choices of $W_0$ will in general lead to different probabilities of the macro-histories. In the Bohmian version, different choices of $W_0$ will lead to different velocity fields such that for typical initial particle configurations (and hence typical worlds compatible with the theory) they will take on different trajectories.%\footnote{The last claim relies on a plausible conjecture: (C) each typical complete particle trajectory $Q(t)$ is compatible with only one initial quantum state under $W$-Bohmian mechanics. If C fails, we may explore a fail-safe. Recall that nomic exactness is compatible with a set of things that has sharp boundaries. We may postulate a disjunctive IPH with a determinate set of  $W(0)$  that are compatible with $Q(t)$. Because of the $W$-guidance equation, the set will be exact.  If the fail-safe is unsatisfactory, we still have the ontological interpretation of $W_0$  that guarantees traceability of IPH.}
 \end{enumerate}
The traceability of $W_0$ is due to the fact  that we have connected the low-entropy macrostate (now represented by $W_0$) to the micro-dynamics (in which  $W_0$ occurs). Hence, $W_0$ plays a dual role at $t_0$ (and only at that time): it is both the microstate and the macrostate. In contrast, the untraceability of $\Gamma_0$ in the classical SPH is due to the fact that  classical equations of motion directly involve only the microstate $X_0$, not $\Gamma_0$. Similarly, the $\mathscr{H}_0$ in the standard wave-function version of QSPH is untraceable because the Schr\"odinger equation directly involves only the wave function, not $\mathscr{H}_0$. There are many changes to $\Gamma_0$ and to $\mathscr{H}_0$ that make no changes whatsoever in typical worlds compatible with those postulates. 

%------------------------------------------------

\section*{Acknowledgement} \y{I thank the editors and two anonymous referees of \textit{The Philosophical Review} for valuable comments. 
I am also grateful for helpful discussions with David Albert, Jacob Barandes, Craig Callender, Eugene Chua, Julianne Chung, Jonathan Cohen, Neil Dewar, Christopher Dorst, Kenny Easwaran, Dorothy Edgington, Sam Elgin, Michael Esfeld, Ronny Fernandez, Sheldon Goldstein, Veronica Gomez, Jeremy Goodman, Patrick Greenough, Alan H\'ajek, Tyler Hildebrand, Christopher Hitchcock, Mario Hubert, Nick Huggett, Tom Imbo, Niels Linnemann, John MacFarlane, Tim Maudlin, Brian McLaughlin, Michael Miller, Kerry McKenzie, Dustin Lazarovici, Barry Loewer,  Oliver Pooley, Diana Raffman, Samuel Rickless, Susanna Rinard, Ezra Rubenstein, Daniel Rubio, Don Rutherford, Mark Sainsbury, Miriam Schoenfield, Charles Sebens, Ayoob Shahmoradi,  Ted Sider, Gila Sher, Elliott Sober, Anncy Thresher, Roderich Tumulka, Yanjing Wang, Eric Watkins,  Isaac Wilhelm, and Eric Winsberg; as well as audiences at CalTech, City College of New York, Columbia University, University of Illinois at Chicago,  UC San Diego,  Delta 13 Logic Workshop, and Harvard Mini-Workshop on the Foundations of Thermodynamics. Thanks especially to Wayne Myrvold and David Wallace for their feedback on my talk at the University of Western Ontario that led to more careful considerations about vagueness and arbitrariness. }

%----------------------------------------------------------------------------------------
%	BIBLIOGRAPHY
%----------------------------------------------------------------------------------------

\bibliography{test}

%----------------------------------------------------------------------------------------

\end{document}